\documentclass[]{aastex631}

\shorttitle{Olivine twin peaks in thermal infrared spectra}
\shortauthors{Kimura et al.}
\graphicspath{{./}{figures/}}

\usepackage{epigraph}

\begin{document}

\title{Do twin spectral peaks of olivine particles in the thermal infrared diagnose their sizes and porosities?}

\correspondingauthor{Hiroshi Kimura}
\email{hiroshi\_kimura@perc.it-chiba.ac.jp}

\author{Hiroshi Kimura}
\affiliation{Planetary Exploration Research Center (PERC), Chiba Institute of Technology, Tsudanuma 2-17-1, Narashino, Chiba 275-0016, Japan}

\author{Johannes Markkanen}
\affiliation{Institut f\"{u}r Geophysik und Extraterrestrische Physik, Technische Universit\"{a}t Braunschweig, Mendelssohnstr. 3, 38106 Braunschweig, Germany}

\author{Ludmilla Kolokolova}
\affiliation{Department of Astronomy, University of Maryland, College Park, MD, 20742, USA}

\author{Martin Hilchenbach}
\affiliation{Max Planck Institute for Solar System Research, Justus-von-Liebig-Weg 3, 37077, G\"{o}ttingen, Germany}

\author{Koji Wada}
\affiliation{Planetary Exploration Research Center (PERC), Chiba Institute of Technology, Tsudanuma 2-17-1, Narashino, Chiba 275-0016, Japan}

\author{Yasumasa Kanada}
\affiliation{Planetary Exploration Research Center (PERC), Chiba Institute of Technology, Tsudanuma 2-17-1, Narashino, Chiba 275-0016, Japan}

\author{Takafumi Matsui}
\affiliation{Planetary Exploration Research Center (PERC), Chiba Institute of Technology, Tsudanuma 2-17-1, Narashino, Chiba 275-0016, Japan}

\begin{abstract}
A well-established constraint on the size of non-porous olivine grains or the porosity of aggregates consisting of small olivine grains from prominent narrow peaks in thermal infrared spectra characteristic of crystalline silicates is reexamined.
To thoroughly investigate thermal infrared peaks, we make theoretical argument for the absorption and scattering of light by non-porous, non-spherical olivine particles, which is followed by numerical verification.
Our study provides perfectly rational explanations of the physics behind the small-particle effect of emission peaks in the framework of classical electrodynamics and convincing evidence of small-particle's emission peaks in the literature.
While resonant absorption excited by surface roughness on the order of submicrometer scales can be identified even for non-porous olivine particles with a radius of $10~\micron$, it makes only a negligible contribution to thermal infrared spectra of the particles.
In contrast, the porosity of non-spherical particles has a significant impact on the strength and wavelength of the peaks, while the resonant absorption excited by an ensemble of small grains takes place at a wavelength different than one expects for surface roughness. 
We finally reaffirm that twin peaks of olivine in thermal infrared spectra of dust particles in astronomical environments are the intrinsic diagnostic characters of submicrometer-sized small grains and their aggregate particles in fluffy and porous configurations.
\end{abstract}


\section{Introduction} \label{sec:intro}

\setlength{\epigraphwidth}{0.53\textwidth}
\begin{epigraphs}
\qitem{There is a class of electromagnetic modes in small particles, called {\it surface modes}, which give rise to interesting---perhaps puzzling at first sight---absorption spectra: small-particle absorption spectra can have features where none exist in the bulk and several features where only a single absorption band exists in the bulk.}{Craig F. Bohren, Donald R. Huffman~\citeyearpar{bohren-huffman1983}\\``Absorption and Scattering of Light by Small Particles''\\  Chapter~12: Surface Modes in Small Particles, p.~325}
\end{epigraphs}

This quotation from one of the classic textbooks on the interaction of electromagnetic waves with small solid particles well summarizes our current understanding of the physics behind the particle-size effect on thermal infrared spectra of dust particles on the whole.
Surface modes are distinct intrinsic conditions of resonant vibrations where the lowest-order mode ($n=1$), also known as the Fr\"{o}hlich mode, plays a major role in infrared spectra of small particles \citep{froehlich1949,huffman1977}.
The concept of the Fr\"{o}hlich mode to interpret an infrared spectral peak of tiny silicate particles was introduced to the astronomical community a half century ago \citep{knacke1968,gilra1972,steyer-et-al1974,huffman1975}.
Since then, a conspicuous narrow emission peak found in the spectrum of solid particles in astronomical environments has been interpreted as the manifestation of Fr\"{o}hlich modes \citep[e.g.,][]{draine1989,anderson2003,jurewicz-et-al2003,wong-et-al2004}.
A thermal infrared spectrum of a dust particle is determined by, apart from the temperature of the particle (i.e., the blackbody spectrum), the wavelength dependence of emissivity $e$.
On the one hand, the emissivity of a single small particle in the Rayleigh scattering regime tends to increase with both the imaginary part of particle polarizability and the size parameter $x$ of the particle defined by $x \equiv 2 \pi a / \lambda$ with the radius $a$ of the particle and the wavelength $\lambda$.
An emission (equivalently absorption) peak appears in a thermal infrared spectrum of a dust particle, if the electric polarizability of the particle is enhanced in a narrow wavelength interval.
The intensification of particle polarizability in a confined spectral range, namely, a prominent narrow peak in thermal infrared spectra of small particles may originate from the excitement of the Fr\"{o}hlich mode \citep{huffman1975,huffman1977,hayashi1984}.
The Fr\"{o}hlich mode, which gives rise to anomalous absorption at the so-called Fr\"{o}hlich frequency owing to the intensification of polarizability, depends not only on the size, the shape, and the structure, but also the composition of the particles \citep{fuchs1974,aronson-emslie1975,huffman1977,koike-et-al2003,koike-et-al2010,ishizuka-et-al2018}.
As a result, we may regard an ensemble of Fr\"{o}hlich frequencies in the thermal infrared spectra of minute dust particles as the unique spectroscopic fingerprint of mineral species embedded in the particles \citep{kimura2014}.
Without doubt, infrared spectroscopic observations of dust particles in cometary comae, the interplanetary medium, and debris disks have long provided the opportunity for us to identify mineral ingredients of the particles by spectral emission peaks \citep{bregman-et-al1987,knacke-et-al1993,reach-et-al2003}.
Mid-infrared spectra of dust particles in planetary systems often contain two major peaks at $\lambda \approx 10.1$ and $11.2~\micron$ in the range of $\lambda = 8$--$13~\micron$ that are attributed to Mg-rich olivine \citep{hanner-et-al1997,wooden-et-al2004,kimura-et-al2008,kimura2014}.
The twin peaks of Mg-rich olivine are the obvious manifestations of the Fr\"{o}hlich modes, because the respective peak lies exactly in the range of wavelengths expected for the Fr\"{o}hlich mode of Mg-rich olivine inclusive of forsterite, the Mg-rich endmember of olivine \citep[cf.][]{huffman1977,mukai-koike1990,sogawa-et-al2006,pitman-et-al2013}.

Laboratory measurements and numerical simulations of thermal emission from silicate dust particles have demonstrated that large particles of $x > 1$ do not exhibit a pronounced narrow silicate peak in their mid-infrared spectra \citep{rose1979,krishnaswamy-donn1979,hanner-et-al1987,mukai-koike1990}.
More precisely, a peak in the infrared spectra of emissivity may rise with particle size, but the sharpness of the peak declines and, in consequence, the absorption spectra of large silicate particles become broad and relatively featureless.
A notable exception is the infrared spectra of extremely porous\footnote{The porosity of fluffy aggregates cannot be uniquely defined, while the definition proposed by \citet{mukai-et-al1992} for fractal aggregates seems to be most commonly adopted in the astronomical community \citep[e.g.,][]{harker-et-al2002,ootsubo-et-al2007,bertini-et-al2009}.} aggregates consisting of submicrometer- or nanometer-sized silicate grains, because they behave like an ensemble of small constituent grains \citep{greenberg-hage1990,okamoto-et-al1994}.
Thermal emission from fluffy aggregates of submicron constituent grains well accounts for the presence of Mg-rich olivine peaks in the infrared spectra of dust particles in cometary comae and debris disks, even though the overall size of the particles far exceeds the wavelength of thermal radiation \citep{kolokolova-et-al2007,kimura-et-al2008}.
The maximum size of fluffy aggregates whose infrared spectra clearly show Mg-rich olivine peaks increases with the porosity of the aggregates, which may be associated with the fractal dimension of the aggregates.
Consequently, there has been consensus about the presence of olivine twin peaks in the infrared spectra as the intrinsic diagnostic characters of fluffy aggregate structures and small constituent grains \citep{hanner-bradley2004}.

An experimental report on the presence of emission peaks in thermal infrared spectra of non-porous, half-millimeter-sized olivine particles by \citet{chornaya-et-al2020} were to cast doubt on the consensus view of olivine twin peaks.
They speculated that the absence of olivine twin peaks in thermal infrared spectra of large particles in the literature is an artifact originating from the assumption of spherical shape for the particles or for constituent grains that make up large porous aggregates in numerical simulations. 
Contrary to their speculation, numerical results with non-spherical particles of cubic shape based on the discrete dipole approximation manifested a broadening of absorption peaks in the infrared spectra for large non-porous particles compared with small ones \citep{ruppin1997}.
In harmony with the numerical simulations, laboratory measurements of thermal emission from irregularly shaped particles of silicates inclusive of olivine revealed the particle-size dependence of silicate emission peaks in their infrared spectra \citep{hunt-logan1972,hunt1976,rose1979}.
There is, therefore, no evidence that the spherical shape of particles is an exception for lack of olivine emission peaks in the infrared spectra at the large particle-size limit.

On the basis of infrared spectroscopic observations, the strengths of olivine emission peaks in the comae of comets vary from one comet to another, but seem to be closely correlated with the orbital semimajor axis of the comets \citep{kolokolova-et-al2007}.
This correlation sounds reasonable from the view point of physical processes in cometary activities, because gas drag due to sublimation of cometary ices selectively carry away small grains and highly porous aggregates owing to their high area-to-mass ratios and hence large grains and relatively compact aggregates tend to accumulate on the surfaces of cometary nuclei \citep{yamamoto-et-al2008}.
On the contrary, if olivine emission peaks were to remain in the infrared spectra of large non-porous particles for whatever reason, a lack of the peaks in the infrared spectra of cometary comae would be attributed to the deficit of olivine in cometary dust.
It is, however, highly unlikely that the abundances of silicate in dust particles greatly differ from comet to comet, because the elemental abundances of dust particles in the comae of comets 1P/Halley, 81P/Wild 2, and 67P/Churyumov-Gerasimenko are all solar and very much alike \citep{kimura-et-al2020a}.
Nevertheless, by considering that articles on small-particle absorption spectra written a half century ago might be nowadays consigned to oblivion, it might be good timing to readdress the Fr\"{o}hlich mode and reaffirm that large non-porous particles fail to retain twin peaks of olivine in their thermal mid-infrared spectra.

We hypothesize that the consensus view of olivine twin peaks is well grounded and any controversial issue could be attributed to misinterpretation of the physics behind the appearance of the peaks.
First, we theoretically make the argument in the framework of classical electrodynamics that non-porous and non-spherical particles cannot produce olivine twin peaks in their thermal infrared spectra, if they are larger than the wavelength of thermal radiation (Sect.~\ref{sec:theoretical}).
The present study focuses on strong narrow twin peaks due to the Fr\"{o}hlich modes of olivine, by putting aside weak peaks originating from surface modes and bulk modes, as well as peaks of other mineral species.
Then, we numerically verify our statements in Sect.~\ref{sec:theoretical} that infrared olivine features of thermal emission from non-porous, non-spherical particles are peaks for small particles, but troughs for large particles (Sect.~\ref{sec:numerical}).
The effects of surface roughness and porosity on the infrared olivine features of thermal radiation are also studied with the benefit of powerful computational tools.
Although our results do no more than approve the long-established constraint on the size and porosity of olivine grains, the obscuration of Fr\"{o}hlich modes by higher-order surface modes is better visualized.
Scientific evidence in Sect.~\ref{sec:theoretical}--\ref{sec:numerical} allows us to reaffirm that prominent olivine twin peaks of non-porous dust particles fades with the size and volume filling factor of the particles, irrespective of their shapes.
Finally, we end with concluding remarks that synthesize a consensus on the interpretation of olivine emission peaks from remote-sensing observational and in-situ experimental points of view (Sect.~\ref{sec:remarks}).

\section{Theoretical argument}
\label{sec:theoretical}

\begin{figure}
\plottwo{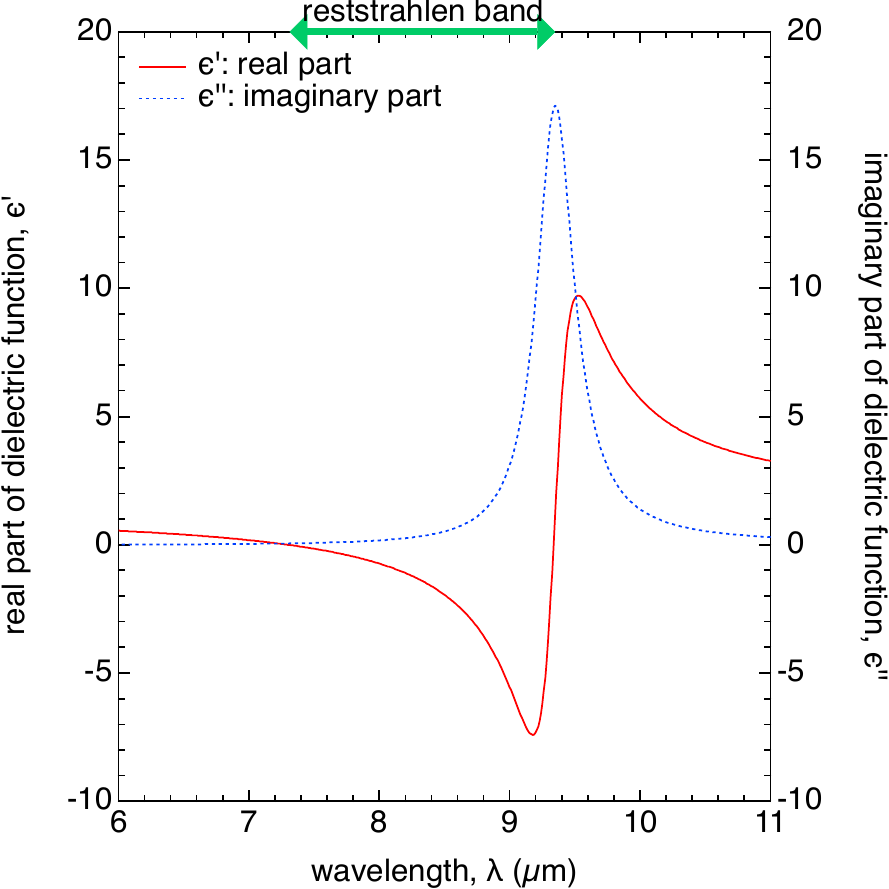}{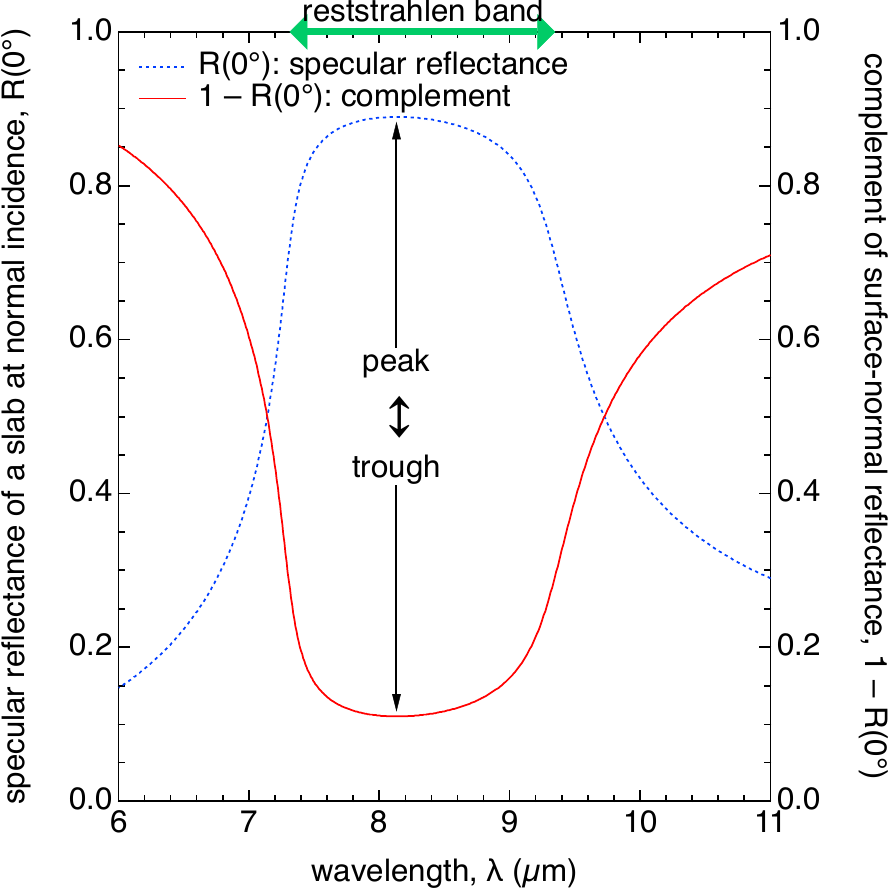}
\caption{Left: the dependence of complex dielectric function $\epsilon = \epsilon' + i \epsilon''$ on wavelength $\lambda$ near the resonance frequency $\omega_0$ calculated from the Lorentz oscillator model.
The resonant frequency $\omega_0$, the collision frequency $\gamma$, and the plasma frequency $\omega_\mathrm{p}$ of the Lorentz oscillator model are set to $\omega_0 =0.107~\micron^{-1}$, $\gamma = 0.004~\micron^{-1}$ and $\omega_\mathrm{p} = 0.0856~\micron^{-1}$, respectively, just as an example \citep{huffman1977}.
Right: the corresponding specular reflectance of a slab at normal incidence, $R(0\degr)$, along the left vertical axis (dotted line) and the quasi-emissivity of $1-R(0\degr)$ for a large non-porous particle along the right vertical axis (solid line), which shows a reflectance peak in the reverse direction, namely, a trough.
The arrows along the top horizontal axes indicate the range of wavelengths that exhibits the reststrahlen band according to this set of parameters for the Lorentz oscillator.
\label{fig:1}}
\end{figure}
The quantity that \citet{chornaya-et-al2020} measured with large (half-millimeter), non-porous olivine particles of irregular shape is apparently not thermal emission from the particles nor the emissivity $e$ of the particles, but the reflectance $R$, which is a fraction of incident irradiance reflected by the particles.
Nonetheless, one might expect that reflectance measurements of large particles would be equivalent to measurements of thermal emission from the particles, if the emissivity of the particles were complementary to the reflectance as $e = 1 - R$.
Indeed, \citet{chornaya-et-al2020} stated that they converted their reflectance spectra into absorption spectra using the equality of $e = 1 - R$, although what actually plotted in their paper is a comparison between their reflectance spectra of large olivine particles and thermal emission spectra of cometary dust, which seemingly escaped notice in the peer-review process.
To be exact, the equality of $e = 1 - R$ is not valid for the reflection of incident waves by a single particle, even if the particle is so large that the transmission of the waves is negligible \citep{rea-welch1963,bohren-huffman1983,lane-et-al2011,hapke2012}.
Instead, the emissivity of a large non-porous particle in the geometrical optics regime approaches the limit $e = 1 - Q_\mathrm{refl}$ where $Q_\mathrm{refl}$ is the reflection efficiency of the particle \citep{bohren-huffman1983,min-et-al2003}.
Namely, the sum of absorption and reflection efficiencies is unity, while the extinction efficiency in the geometrical optics limit is equal to two---its maximum value, because the diffraction efficiency is unity \citep{nussenzveig-wiscombe1980,bohren-huffman1983,min-et-al2003}.
The transmission efficiency of large particles approaches zero, because all the photons that penetrate a sufficiently large non-porous particle are absorbed, except for non-absorbing particles, whereas the other photons are reflected or diffracted \citep{bohren-huffman1983,hapke2012}.
The supposition of  $e = 1 - R$ is, nevertheless, not a fatal flaw in the geometrical optics limit, provided that the reflection efficiency $Q_\mathrm{refl}$ is crudely proportional to the specular reflectance $R(0\degr)$ at normal incidence \citep{schlick1994,hapke2012}.
It is worthwhile noting that the infrared spectra of reflectance exhibit a broad peak in a so-called reststrahlen band, where the real part $\epsilon'$ of complex dielectric function $\epsilon = \epsilon' + i \epsilon''$ takes negative values \citep{huffman1977,bohren-huffman1983,hapke2012}.
High reflectance in a reststrahlen band is intuitively understood by inserting $\epsilon' = - \epsilon_\mathrm{r}$ with $\epsilon_\mathrm{r} > 0$ into the Fresnel's equation of $R(0\degr) = \left|{\left({1-\sqrt{\epsilon}}\right) / \left({1+\sqrt{\epsilon}}\right)}\right|^2$, from which we get $R(0\degr) \simeq \left({1+\epsilon_\mathrm{r}}\right)/\left({1+\epsilon_\mathrm{r}}\right) = 1$ for $\epsilon'' \ll \epsilon_\mathrm{r}$.
Therefore, the presence of peaks in infrared reflectance spectra of large olivine particles comes as no surprise, unless the real part of complex dielectric function is positive throughout the whole wavelength range.
As a more realistic example, we may adopt the Lorentz model of one oscillator with the resonant, the collision, and the plasma frequencies, which are denoted by $\omega_0$, $\gamma$, and $\omega_\mathrm{p}$, respectively \citep{huffman1977}.
In the Lorentz model for the oscillatory motion of bound electrons in a solid under their interactions with electromagnetic waves, the resonant frequency $\omega_0$, the collision frequency $\gamma$, and the plasma frequency $\omega_\mathrm{p}$ are associated with restitution (Hooke's law), friction, and Coulomb forces, respectively \citep{bohren-huffman1983,jackson1998}.
The solution to the equation of motion gives the expression for the dependence of complex dielectric function on the frequency of the waves through the spectral variations of $\omega_0$, $\gamma$, and $\omega_\mathrm{p}$.
The left panel of Fig.~\ref{fig:1} depicts how the real part (solid line) and the imaginary part (dotted line) of complex dielectric function change with wavelength near the resonance at $\lambda = 1/ \omega_0 \approx 9.35~\micron$ for $\omega_0 =0.107~\micron^{-1}$, $\gamma = 0.004~\micron^{-1}$, and $\omega_\mathrm{p} = 0.0856~\micron^{-1}$ \citep{huffman1977}.
The right panel of Fig.~\ref{fig:1} illustrates how the real part of complex dielectric function given in the left panel affects the specular reflectance at normal incidence (dotted line).
Because only the real part of complex dielectric function matters, a peak in a reststrahlen band steadily manifests in the infrared spectra of reflectance, no matter what size of particles is concerned \citep{salisbury-wald1992,salisbury-et-al1994}.
We should, however, emphasize that a spectral peak of reflectance $R$ in a reststrahlen band is transformed into a trough (i.e., the reversal of a peak) in the spectral variation of $1 - R$, the proof of which is nothing more than arithmetic exercise (see the solid line in the right panel of Fig.~\ref{fig:1}).
Consequently, even a thought experiment based on the supposition of $e = 1 - R$ proves that large olivine particles do not necessarily show emission peaks in their thermal infrared spectra, but troughs, namely, peaks in the reverse direction.

A pronounced narrow absorption peak in the infrared spectra of thermal emission from dust particles is known to appear near the Fr\"{o}hlich frequency in the range of $\epsilon' \la 0$, provided that the particles are much smaller than the wavelength of thermal radiation \citep{bohren-huffman1983}.
The lowest-order surface mode---the Fr\"{o}hlich mode produces uniform electric fields inside the particle, although the terminology ``surface modes'' is associated with electric fields confined near the surface of the particle in the higher-order modes.
It is the uniformity of electric fields inside the particle that induces oscillations of all the molecules in phase and, in consequence, resonant absorption, which appears as a narrow peak in the thermal infrared spectrum of the particle.
The Fr\"{o}hlich frequency depends on the size, shape, structure, and composition of particles, because the Fr\"{o}hlich mode is linked to the polarizability of the particles \citep{ruppin-englman1970,huffman1977}.
Infrared absorption spectra of non-porous particles with spherical and cubic shapes calculated from Mie theory and the discrete dipole approximation demonstrated that the Fr\"{o}hlich frequency decreases and the peak broadens with the size of the particles \citep{bohren-huffman1983,ruppin1997}.
We should emphasize that the broadening of peaks in the infrared spectra is a general trend for large particles of arbitrary shape, originating from the excitement of higher-order surface modes \citep{bohren-huffman1983}.
In addition, the Fr\"{o}hlich frequency usually decreases with the size of particles and, eventually, moves out of the wavelength range where the real part of complex dielectric function is negative.
In other words, the concept of Fr\"{o}hlich frequency cannot be applied to the absorption of light by a sufficiently large particle, because negative values of $\epsilon'$ are no longer situated in a proper range of wavelengths for such a particle.
Unless there is a specific circumstance that prevents the excitement of higher-order surface modes and the shift of the Fr\"{o}hlich frequency, large dust particles cannot meet the necessary condition for the development of a prominent narrow absorption peak no matter what shape and structure they have.
It is, therefore, inevitable that anomalous absorption at the Fr\"{o}hlich frequency is characteristic of, if non-porous, small particles in the Rayleigh scattering regime, because they cannot excite higher-order surface modes.

\section{Numerical verification}
\label{sec:numerical}

\begin{figure}
\plottwo{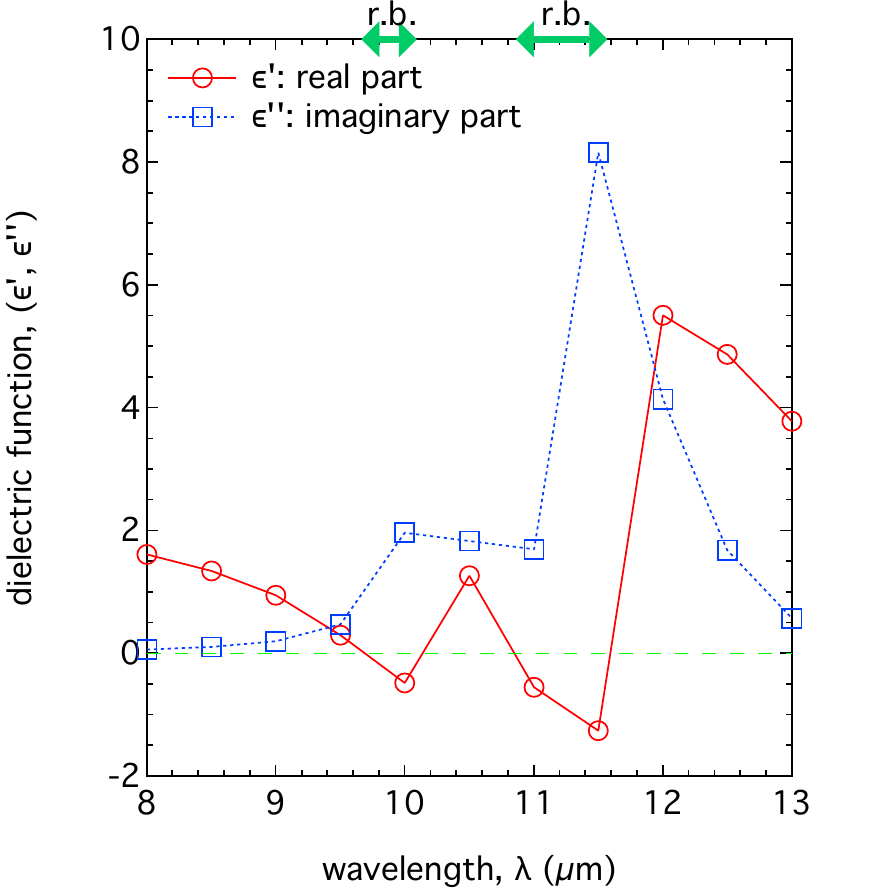}{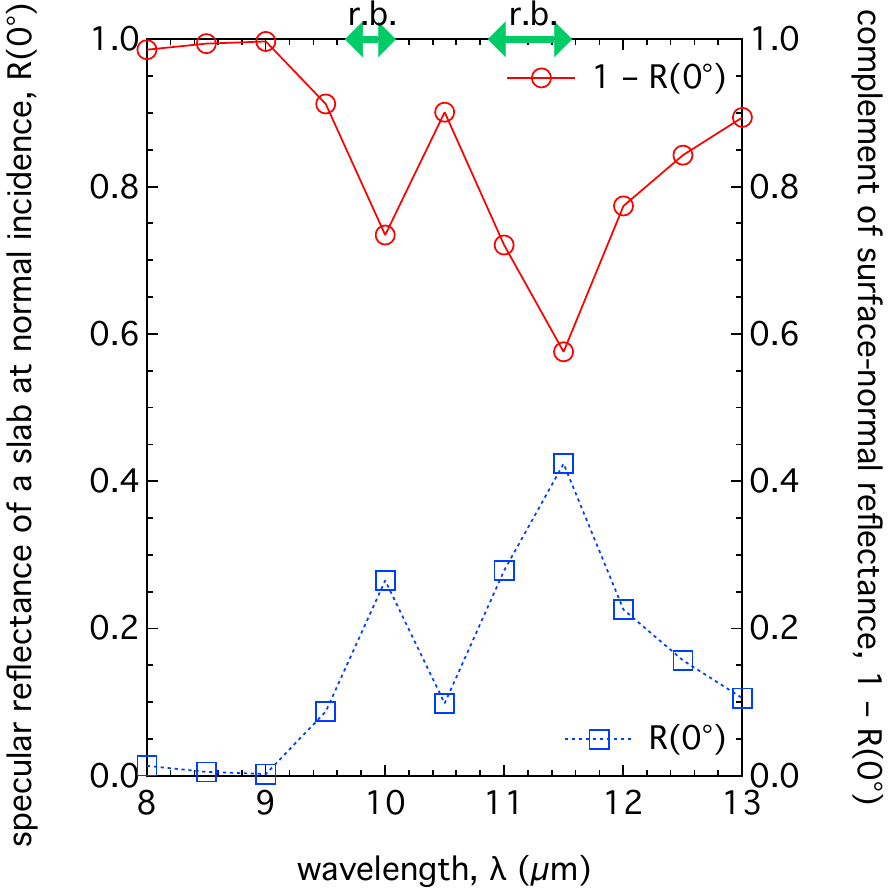}
\caption{Left: a spectral variation in the real part $\epsilon'$ (open circles) and imaginary part $\epsilon''$ (open squares) of complex dielectric function $\epsilon = \epsilon' + i \epsilon''$ for Mg-rich olivine that are used for computations of absorption efficiency and geometric albedo.
The complex refractive indices $m$ of olivine determined by \citet{mukai-koike1990} are used to calculate the complex dielectric functions $\epsilon$ with the help of the relation $\epsilon = m^2$.
Right: the specular reflectance of a slab at normal incidence, $R(0\degr)$, (open squares) calculated from the $\epsilon$ values of Mg-rich olivine and the complement of the surface-normal reflectance, $1-R(0\degr)$ (open circles).
The arrows along the top horizontal axes indicate the range of frequencies that exhibits the reststrahlen band (r.b.) as expected from the spectral variation in the real part $\epsilon'$ of complex dielectric function for Mg-rich olivine.
\label{fig:2}}
\end{figure}
Subsequent to the theoretical argument, we aim to quantitatively demonstrate that olivine emission peaks turn into troughs, but reflection peaks in reststrahlen bands remain as the size of non-porous, non-spherical particles increases.
We have to, however, admit that the following results do not necessarily provide new insights into the effect of particle size on the infrared spectra of non-porous, non-spherical particles, since they do not differ qualitatively from previous studies \citep[e.g.,][]{ruppin1997}.
The absorption efficiency $Q_\mathrm{abs}$ ($=e$, according to the Kirchhoff's law) and geometric albedo $A_\mathrm{p}$ of a particle may be considered as the non-dimensional quantities that describe the emissivity and reflectance of the particle \citep{hapke2012,beck-et-al2021}.
Accordingly, we compute their wavelength dependences for randomly oriented, non-spherical particles composed of Mg-rich olivine not only with various sizes, but also with various degrees of surface roughness and porosity.
The absorption efficiency of an arbitrary shaped dust particle is hereafter defined as the absorption cross section $C_\mathrm{abs}$ divided by the geometric cross section $G$ of the particle, namely, $Q_\mathrm{abs} \equiv C_\mathrm{abs} / G$ \citep{peake1959,bohren-huffman1983}.
The geometric albedo of an arbitrary shaped dust particle is described as $A_\mathrm{p} = \pi S_{11}(180\degr) /\left({k^2  G }\right)$ where $S_{11}(180\degr)$ is the $(1,1)$ element of the $4 \times 4$ Mueller matrix at a scattering angle of $180\degr$ and $k = 2 \pi / \lambda$ is the wavenumber \citep{hanner-et-al1981,kimura-et-al2003,kimura-et-al2006}.
The shape of dust particles in the Rayleigh scattering limit is known to affect the wavelength and strength of olivine absorption peaks, as expected for the shape dependence of the Fr\"{o}hlich mode \citep[e.g.,][]{huffman1977,koike-et-al2010}.
Nevertheless, we represent non-porous, non-spherical particles by regular tetrahedra for simplicity, because the same results are expected for any other convex shapes in the geometrical optics limit \citep[e.g.,][]{min-et-al2003}.
It has become common practice to describe the effective radius of nonspherical particles using the volume-equivalent radius $a_\mathrm{V} \equiv \left[{3V / \left({4 \pi}\right)}\right]^{1/3}$ of a sphere with volume $V$.
The volume-equivalent radius of a regular tetrahedron is associated with the edge length $l$ of the tetrahedron as $l = 2 \left({2 \pi^2}\right)^{1/6} a_\mathrm{V}$ and the geometric cross section $G = \sqrt{3} \left({2 \pi^2}\right)^{1/3} a_\mathrm{V}^2$, because the volume $V$ and geometric cross section $G$ of a regular tetrahedron with a edge length of $l$ are $V = \sqrt{2} l^3 / 12$ and $G = \sqrt{3} l^2 / 4$ when averaged over random orientations.
We take the complex refractive indices $m = \sqrt{\epsilon}$ of Mg-rich olivine from \citet{mukai-koike1990} and the real and imaginary parts of complex dielectric function computed with the complex refractive indices are plotted in the left panel of Fig.~\ref{fig:2}.
The right panel of Fig.~\ref{fig:2} depicts the reststrahlen bands of Mg-rich olivine in the specular reflectance of a slab at normal incidence, $R(0\degr)$ (open squares) calculated from the $\epsilon$ values and the complement of the surface-normal reflectance, $1-R(0\degr)$ (open circles).
It is worthwhile noting that the Fr\"{o}hlich modes and reststrahlen bands are expected to appear in the range of wavelengths where the real part $\epsilon'$ of complex dielectric function is negative, the range of which is indicated by arrows on the top horizontal axes of Fig.~\ref{fig:2} \citep{bohren-huffman1983}.
We should mention that the two negative $\epsilon'$ regions are entirely consistent with other experimental data on the dielectric functions of olivine and forsterite measured independently \citep{huffman1977,sogawa-et-al2006,pitman-et-al2013}.

\subsection{The effect of particle size}

\begin{figure}
\plotone{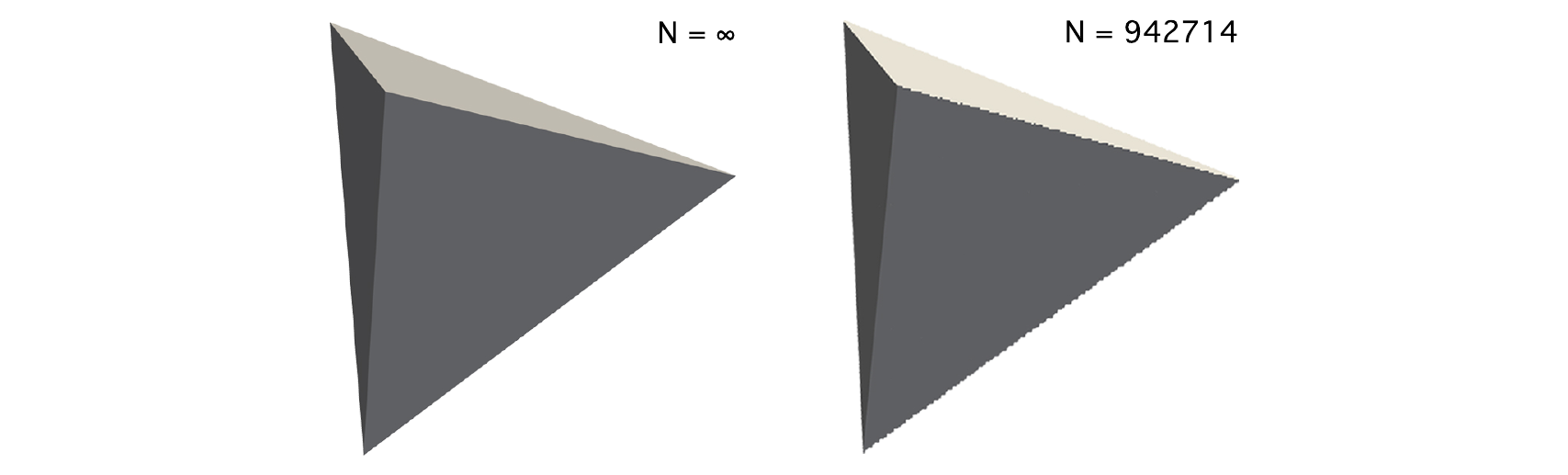}\\
\epsscale{1.0}
\plottwo{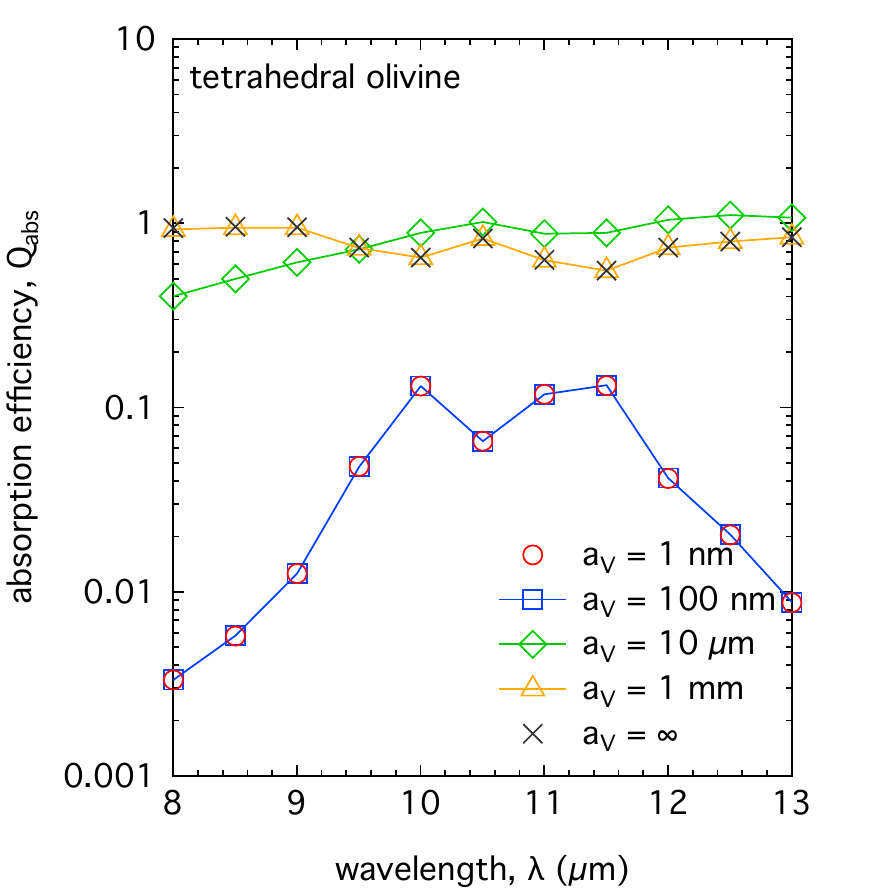}{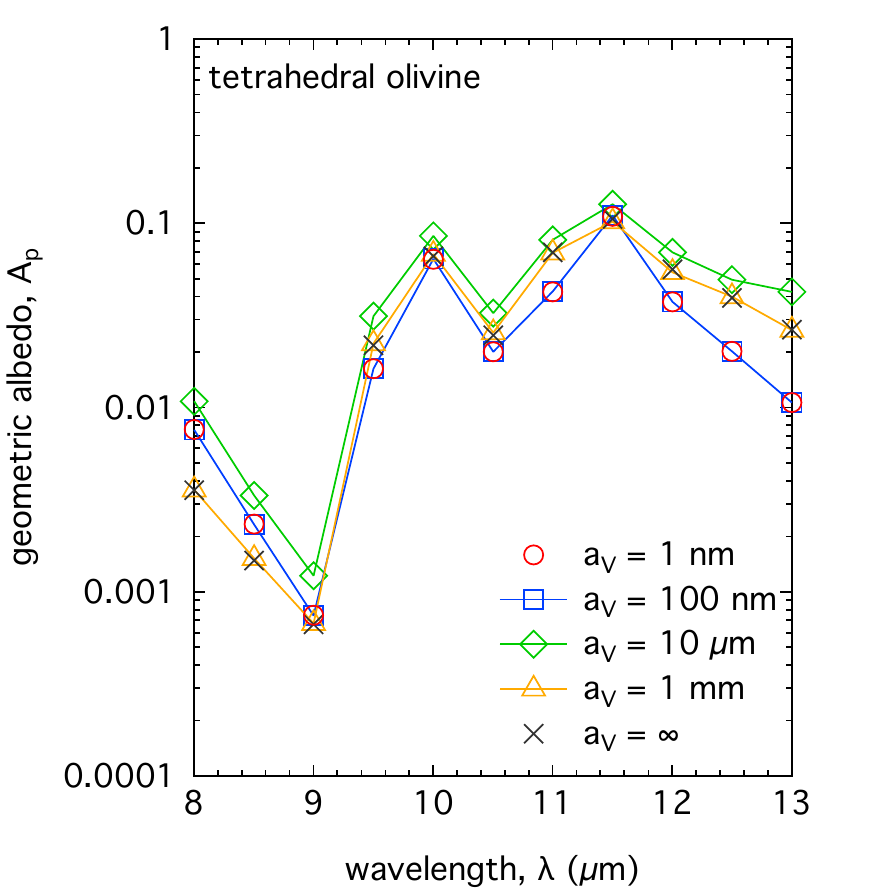}
\caption{Top: Non-porous regular tetrahedra with perfectly smooth surfaces (left) and the packed configuration of $N=942714$ dipoles in a regular tetrahedral shape (right) used for computations of absorption efficiency $Q_\mathrm{abs}$ and geometric albedo $A_\mathrm{p}$.
Bottom: the absorption efficiency $Q_\mathrm{abs}$ (left) and geometric albedo $A_\mathrm{p}$ (right) of tetrahedral olivine particles as a function of wavelength $\lambda$ in the range of $\lambda = 8$--$13~\micron$.
The radius $a_\mathrm{V}$ of volume-equivalent spheres chosen for computations are $a_\mathrm{V} = 1~\mathrm{nm}$ (open circles), $100~\mathrm{nm}$ (open squares), $10~\micron$ (open diamonds), $ 1~\mathrm{mm}$ (open triangles), and $\infty$ (crosses). 
Numerical values of absorption efficiency for $a_\mathrm{V} = 1~\mathrm{nm}$ are multiplied by $10^2$ and those of geometric albedo for $a_\mathrm{V} = 1$ and $ 100~\mathrm{nm}$ by $10^{12}$ and $10^4$, respectively.
\label{fig:3}}
\end{figure}
To our best knowledge, an analytical solution to Maxwell's equations with the boundary condition of a regular tetrahedron has not been found to date.
Therefore, to study the size dependence of absorption efficiency and geometric albedo for regular tetrahedral particles with smooth surfaces, we seek numerical solutions by using the discrete dipole approximation (DDA) for small and intermediate-sized particles with $a_\mathrm{V} \le 10~\micron$ and the ray tracing method (RTM) for large particles of $a_\mathrm{V} = 1~\mathrm{mm}$ \citep{draine-flatau1994,lindqvist-et-al2018}.
Figure~\ref{fig:3} on the left top shows a non-porous regular tetrahedron with perfectly smooth surfaces whose shape and structure are imitated by a large number of dipoles as shown in the right top illustration.
In the DDA computations, we utilize the DDSCAT code (version 7.3) with the lattice dispersion relation proposed by \citet{gutkowiczkrusin-draine2004} for the prescription of dipole polarizability \citep{draine-flatau2008}.
The shape of a regular tetrahedral particle is configured with a sufficiently large number $N$ of scattering units, called dipoles, located in a cubic lattice whose edge length $d$ must fulfill the condition $\left|{m}\right| k d \la 1.0$.
To ensure smooth flat surfaces on the four triangular facets of regular tetrahedral particles, $N=942714$ dipoles are used for the dipole configuration, which gives $d =0.16~\micron$ and $\left|{m}\right| k d < 0.26$ (see the right top illustration of Fig.~\ref{fig:3}).

In the bottom panel of Fig.~\ref{fig:3}, we plot our numerical results on the values of $Q_\mathrm{abs}$ (left) and $A_\mathrm{p}$ (right) as a function of wavelength $\lambda$ for tetrahedral olivine particles with $a_\mathrm{V} = 1~\mathrm{nm}$ (open circles), $ 100~\mathrm{nm}$ (open squares) $10~\micron$ (open diamonds), and $1~\mathrm{mm}$ (open triangles).
Numerical values of absorption efficiency for $a_\mathrm{V} = 1$ and geometric albedo for $a_\mathrm{V} = 1$ and $ 100~\mathrm{nm}$ are multiplied by $10^2$, $10^{12}$, and $10^4$, respectively, because the values are too small to be shown in the figure along with the results for larger particles.
For comparison, we consider a bulk of olivine with the asymptotic values of absorption efficiency and geometric albedo at the large-size limit of $a_\mathrm{V} = \infty$ (crosses), which are given by $Q_\mathrm{abs} = 1 - Q_\mathrm{refl}$ and $A_\mathrm{p} = R(0\degr)/4$ \citep{bohren-huffman1983,hapke2012,chang-et-al2005}.
By computing the values of $Q_\mathrm{abs}$ and $A_\mathrm{p}$ with the help of the DDA for $a_\mathrm{V} \le 10~\micron$  and the RTM for $a_\mathrm{V} = 1~\mathrm{mm}$, we find that the values for $a_\mathrm{V} = 1~\mathrm{mm}$ are virtually indistinguishable from those for $a_\mathrm{V} = \infty$.
The absorption efficiency $Q_\mathrm{abs}$ of olivine tetrahedron particles shows a spectral variation with twin peaks for small particles of $a_\mathrm{V} \le 100~\mathrm{nm}$, but with twin troughs for large particles of $a_\mathrm{V} \ge 1~\mathrm{mm}$.
The former is characteristic of the Fr\"{o}hlich modes in the Rayleigh scattering regime (i.e., $a_\mathrm{V} \ll \lambda$) and the latter of higher-order surface modes in the geometrical optics regime (i.e., $a_\mathrm{V} \gg \lambda$).
While one negative $\epsilon'$ region may produce multiple absorption peaks depending on the shape and structure of the particles, it is unclear how many peaks should appear for small regular tetrahedra in this wavelength range \citep{fuchs1974,huffman1977}.
Although a detailed study on the fine structure of every peak is beyond the scope of this paper, our results with $a_\mathrm{V} = 100~\mathrm{nm}$ are entirely consistent with numerical results of \citet{yanamandrafisher-hanner1999} for submicometer-sized tetrahedral particles of forsterite.
Intermediate-sized particles with $a_\mathrm{V} = 10~\micron$, which is comparable to the wavelength (i.e., $a_\mathrm{V} \approx \lambda$), manifest a relatively featureless, transitional profile in the spectra of their absorption efficiencies.
More precisely, broader faint twin peaks at longer wavelengths appear in the spectra of particles with $a_\mathrm{V} = 10~\micron$, compared to Rayleigh scatterers, owing to a decrease in the Fr\"{o}hlich frequencies and excitements of higher-order surface modes involved in an increasing size of the particles.

In contrast to the wavelength dependence of $Q_\mathrm{abs}$, the geometric albedo $A_\mathrm{p}$ of olivine tetrahedron particles is characterized, irrespective of the particle size, by twin peaks of reststrahlen bands.
The absolute values for small particles with $a_\mathrm{V} = 1$ and $ 100~\mathrm{nm}$, respectively, are twelve and four orders of magnitude below the values for intermediate to large particles with $a_\mathrm{V} \ge 10~\micron$.
In the geometrical optics limit, twin peaks of geometric albedo in reststrahlen bands are paired with twin troughs of absorption efficiency, whereas this is not the case in the Rayleigh scattering limit.
Consequently, our numerical results verify that prominent narrow silicate peaks disappear in emission (left) but remain in reflection (right) as the radius of non-porous, non-spherical olivine particles increases.

\subsection{The effect of surface roughness}

\subsubsection{The concentration of surface humps}

\begin{figure}
\plotone{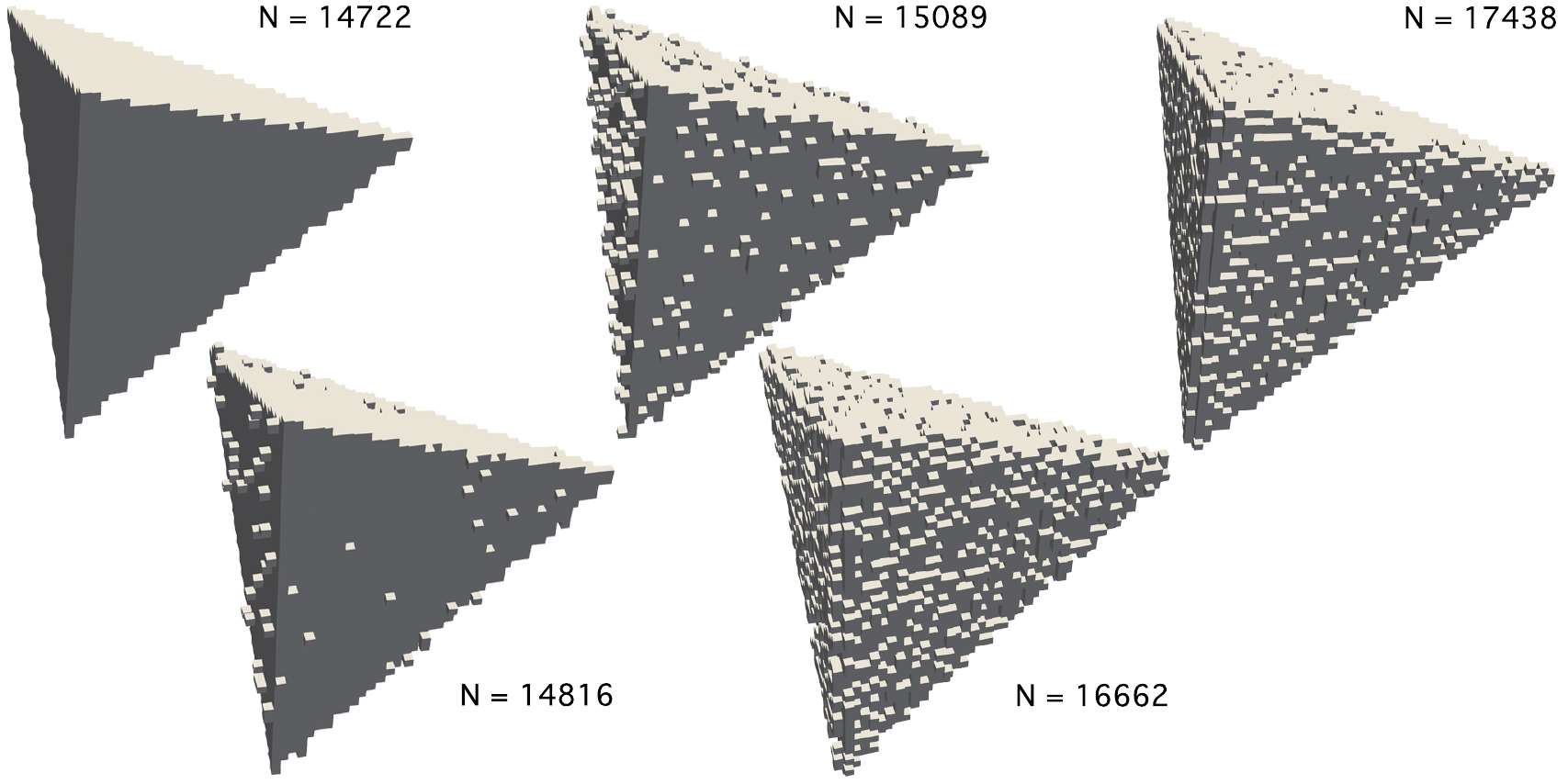}
\plottwo{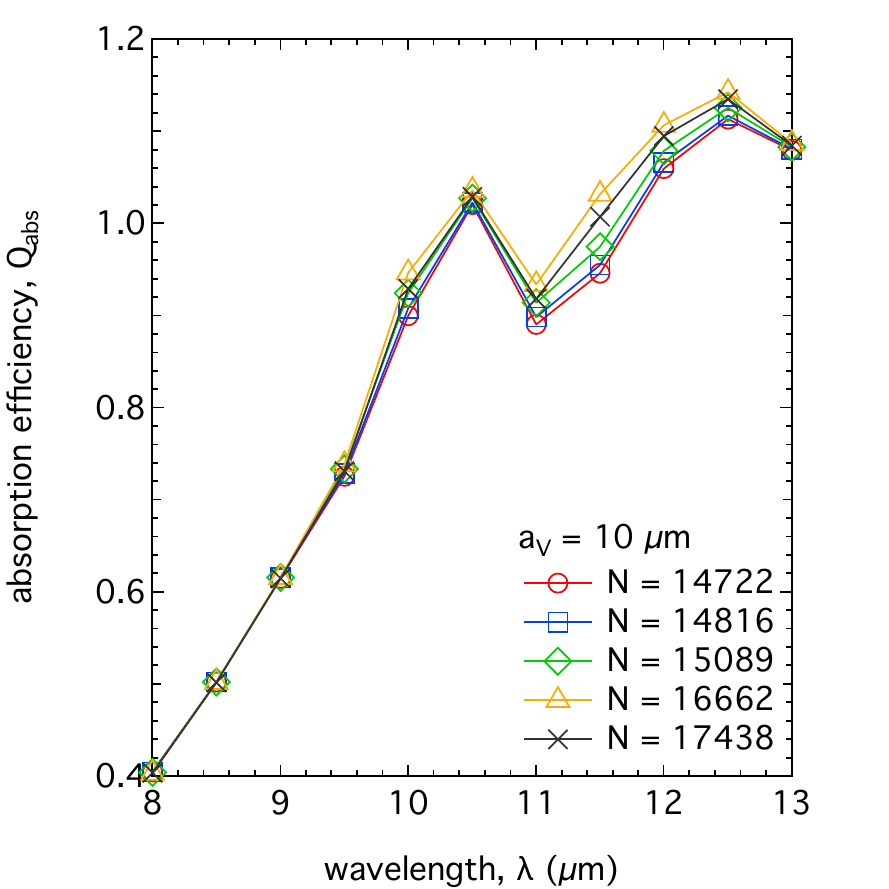}{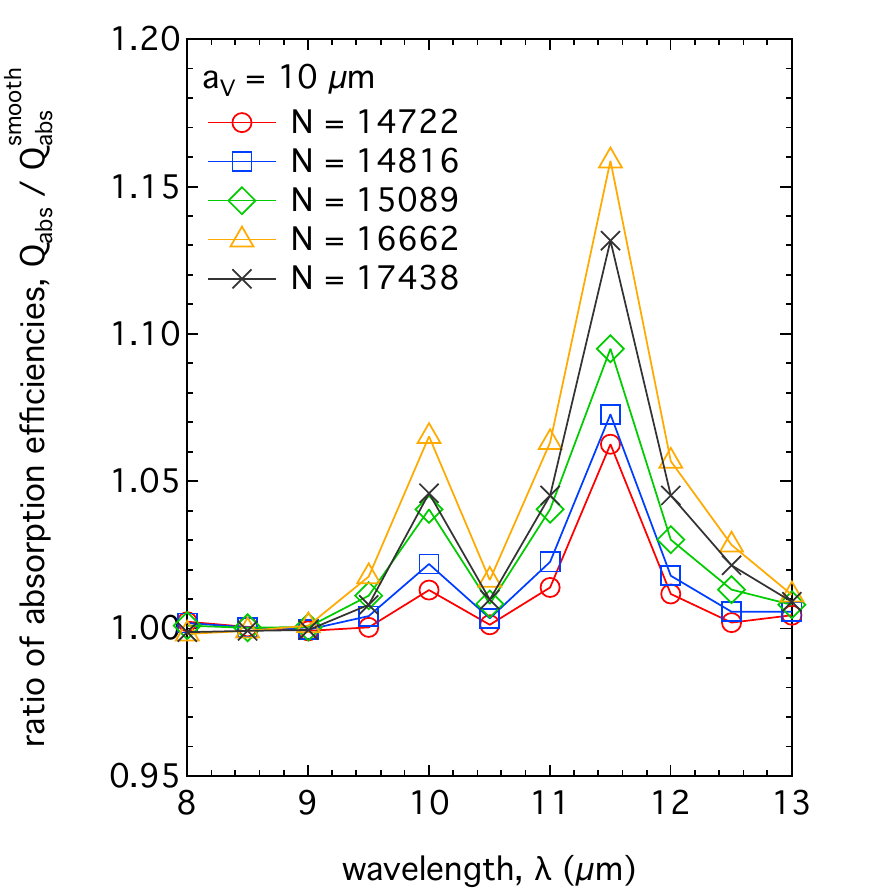}
\caption{Top: five different configurations of $N$ dipoles in a regular tetrahedron with rugged surfaces that represent non-porous particles with an intermediate size of $a_\mathrm{V} = 10~\micron$.
The surface roughness on the order of $d \approx 0.62$--$0.66~\micron$ is simulated with randomly dispersed dipoles on the surfaces of regular tetrahedral particles ($N=14722$--$17438$).
Bottom left: the absorption efficiency $Q_\mathrm{abs}$ of tetrahedral olivine particles with rough surfaces for an intermediate size of $a_\mathrm{V} = 10~\micron$ as a function of wavelength $\lambda$ in the range of $\lambda = 8$--$13~\micron$;
open circles: $N=14722$; open squares: $N=14816$; open diamonds: $N=15089$; open triangles: $N=16662$; crosses: $N=17438$.
Bottom right: The absorption efficiency $Q_\mathrm{abs}$ of tetrahedral olivine particles with rough surfaces normalized to the absorption efficiency $Q_\mathrm{abs}^\mathrm{smooth}$ of a tetrahedral olivine particle with smooth surfaces (i.e., $N=942714$).
\label{fig:4}}
\end{figure}
We examine the effect of surface roughness on the spectral variation of absorption efficiency first by increasing the lattice spacing $d$ on the assumption that the surface roughness is on the order of $d$ and subsequently by adding dipoles sparsely on the surfaces.
The top panel of Fig.~\ref{fig:4} illustrates the configurations of $N=14722$--$17438$ dipoles in intermediate-sized ($a_\mathrm{V} = 10~\micron$) regular tetrahedral particles with surface roughness of $d \approx 0.63$--$0.66~\micron$ ($\left|{m}\right| k d \la 1.0$).
Since every triangular facet of the particles resembles the circumstance that submicrometer-sized grains with diameter $d \approx 0.62$--$0.66~\micron$ are dispersed on a flat surface, one might expect that infrared spectra of the particles exhibit silicate emission peaks of the grains in the Rayleigh scattering regime.

Contrary to the faint expectations, we find no visible effect of surface roughness on the infrared spectra of regular tetrahedral particles with rugged facets, when the calculated values of $Q_\mathrm{abs}$ for $a_\mathrm{V} = 10~\micron$ are plotted on the same scale as Fig.~\ref{fig:3}. 
The visibility of the effect is slightly improved, if the vertical scale is greatly enlarged as shown in the bottom left panel of Fig.~\ref{fig:4}, although the values look very much alike at first glance.
With the intention of visualizing the Fr\"{o}hlich modes, the bottom right panel of Fig.~\ref{fig:4} depicts the values of $Q_\mathrm{abs}$ for $N=14722$--$17438$, normalized to the value for $N = 942714$, which is denoted as $Q_\mathrm{abs}^\mathrm{smooth}$.
Remarkably, a subtle distinction between various amounts of humps on the surfaces is no doubt identified in the bottom right panel of Fig.~\ref{fig:4}, namely, the Fr\"{o}hlich modes of tiny surface roughness become visible.
The Fr\"{o}hlich frequencies for the intermediate-sized regular tetrahedral particles with rugged facets are comparable to those for small ($a_\mathrm{V} = 100~\mathrm{nm}$) particles with smooth facets.
The ratios of absorption efficiencies monotonically increases with the number $\Delta N$ of additional dipoles (i.e., submicrometer-sized humps) in the range of $N=14722$--$16662$ and then decreases with $\Delta N$ at $N>16662$.
Here we would like to mention that the surface fraction $f_\mathrm{hump}$ of submicrometer-sized humps straightforwardly explains this dependence of the $Q_\mathrm{abs}/Q_\mathrm{abs}^\mathrm{smooth}$ ratio on $\Delta N$. 
As the number $\Delta N$ of additional dipoles varies from $\Delta N =0$ at $N=14722$ to $\Delta N = 2716$ at $N=17438$, the surface fraction $f_\mathrm{hump}$ of humps monotonically increases and reaches $f_\mathrm{hump} \approx 0.5$ at $N=16662$.
If the division of the surface without humps is regarded as dips, then the surface appears to be occupied by submicrometer-sized humps and dips half each at $f_\mathrm{hump} = 0.5$.
Because at $f_\mathrm{hump} > 0.5$, dips are smeared with humps, which elevates the surface smoothness, the Fr\"{o}hlich modes of an intermediate-sized regular tetrahedral particle with rough surfaces are most visible at $f_\mathrm{hump} \approx 0.5$.
However, the Fr\"{o}hlich modes of tiny surface roughness play only a minor role in the wavelength dependence of $Q_\mathrm{abs}$ for intermediate-sized particles with $a_\mathrm{V} = 10~\micron$, as shown in the bottom left panels of Figs.~\ref{fig:3} and \ref{fig:4} where the overall profiles of the spectra are characterized by faint twin peaks at longer wavelengths formed by higher-order surface modes.

\subsubsection{The size of surface humps}

\begin{figure}
\plottwo{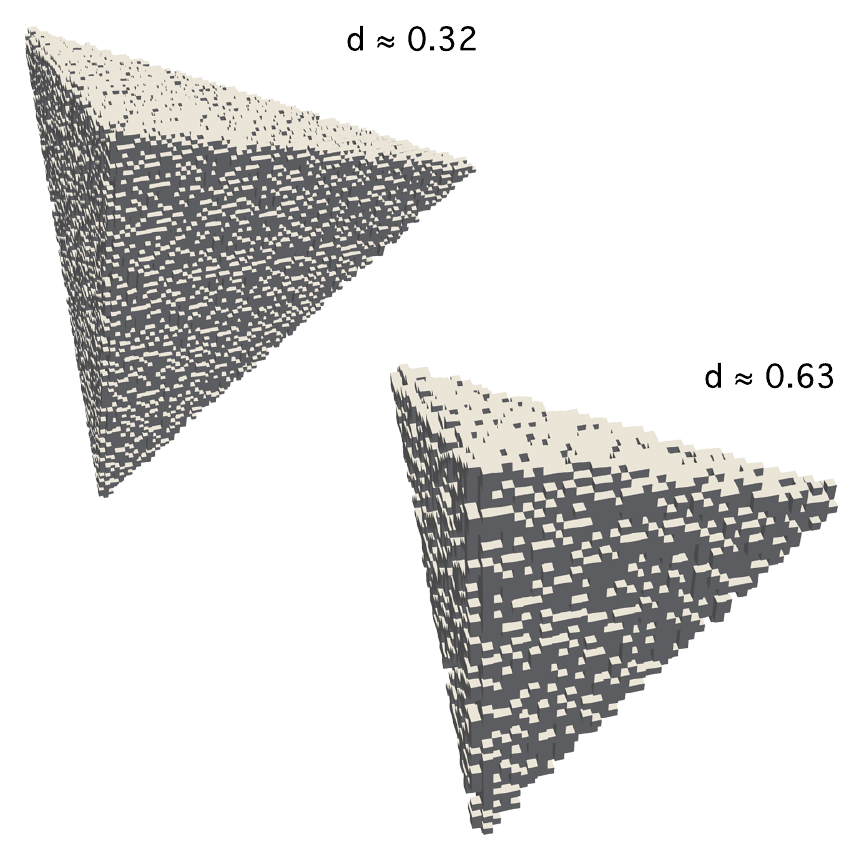}{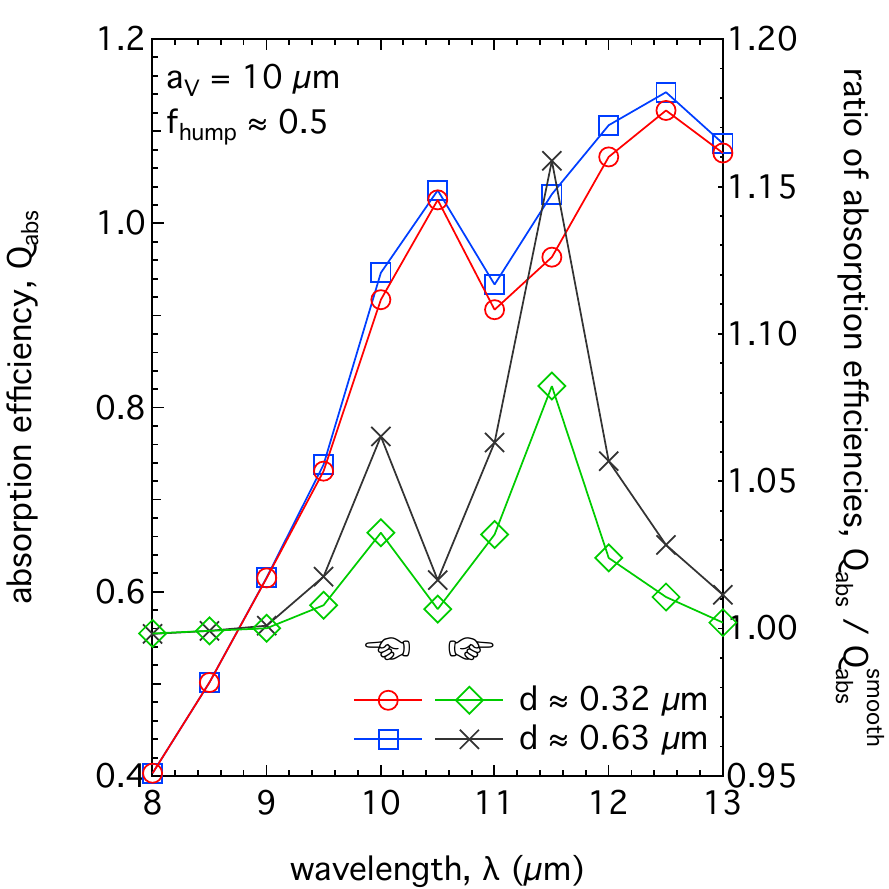}
\caption{Left: two different configurations of $N$ dipoles in a regular tetrahedron with rugged surfaces that represents a non-porous particle with an intermediate size of $a_\mathrm{V} = 10~\micron$.
The surface roughnesses on the order of $d \approx 0.32$ and $0.63~\micron$ are simulated with randomly dispersed dipoles on the surfaces of regular tetrahedral particles with $N=125578$ and $16662$, respectively, both of which set the surface fraction $f_\mathrm{hump}$ of humps to $f_\mathrm{hump} \approx 0.5$.
Right: the absorption efficiency $Q_\mathrm{abs}$ of regular tetrahedral olivine particles with rough surfaces for an intermediate size of $a_\mathrm{V} = 10~\micron$ (left axis) as a function of wavelength $\lambda$ in the range of $\lambda = 8$--$13~\micron$;
open circles: $d \approx 0.32~\micron$; open squares: $d \approx 0.63~\micron$.
Also plotted (right axis) is the wavelength dependence of absorption efficiency normalized to the absorption efficiency $Q_\mathrm{abs}^\mathrm{smooth}$ of a regular tetrahedral olivine particle with smooth surfaces (i.e., $N=942714$): open diamonds: $d \approx 0.32~\micron$ ; crosses: $d \approx 0.63~\micron$.
\label{fig:5}}
\end{figure}
We further investigate whether a larger number of smaller humps on the facets at $f_\mathrm{hump} \approx 0.5$ would enhance the contribution of surface roughness to the Fr\"{o}hlich modes.
Accordingly, we reduce the size of dipole spacing by half ($d \approx 0.32~\micron$) and increase the number of humps per unit area under the condition of $f_\mathrm{hump} \approx 0.5$ as shown in the left panel of Fig.~\ref{fig:5}.
The right panel of Fig.~\ref{fig:5} shows the wavelength dependences of absorption efficiency $Q_\mathrm{abs}$ of tetrahedral olivine particles with rough surfaces of $d \approx 0.32$ (open circles) and $d \approx 0.63~\micron$ (open squares).
Also plotted are the absorption efficiency normalized to the absorption efficiency $Q_\mathrm{abs}^\mathrm{smooth}$ of a tetrahedral olivine particle with smooth surfaces (i.e., $N=942714$) for the particles with rigged surfaces of $d \approx 0.32$ (open diamonds) and $d \approx 0.63~\micron$ (crosses).
The numerical results indicate that the size and number of humps does not distinctly affect the thermal infrared spectra of non-porous olivine particles.
It is, therefore, safe to rule out the possibility that surface roughness on the order of submicrometer scales produce narrow peaks in the infrared spectra of thermal emission from large silicate particles of $a_\mathrm{V} \ga 10~\micron$.

\subsection{The effect of particle porosity}

\subsubsection{A single large closed cavity}

\begin{figure}
\plotone{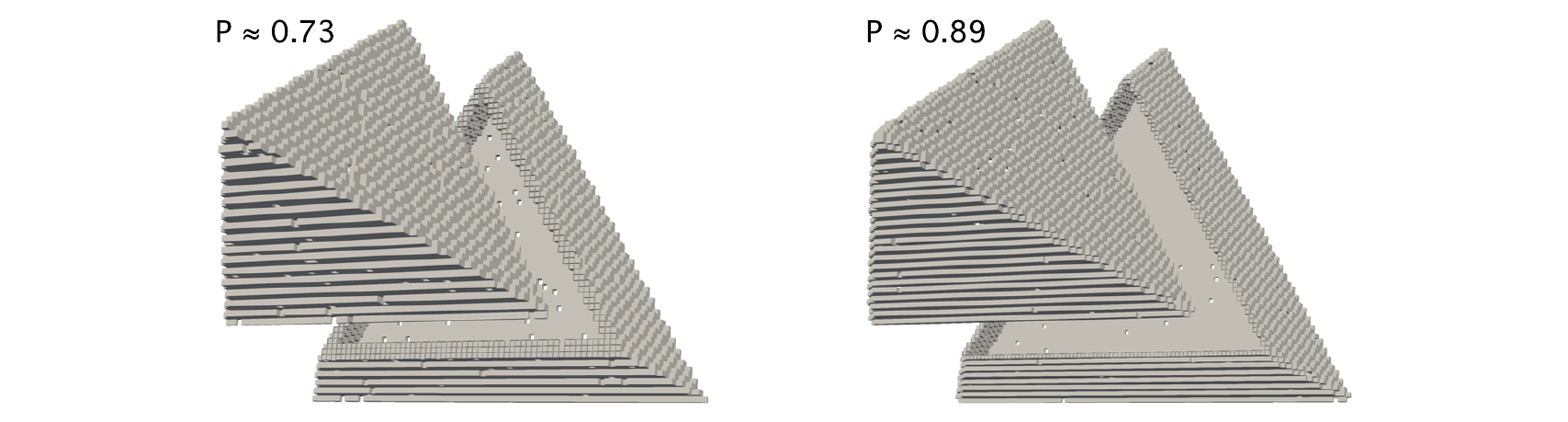}\\
\plottwo{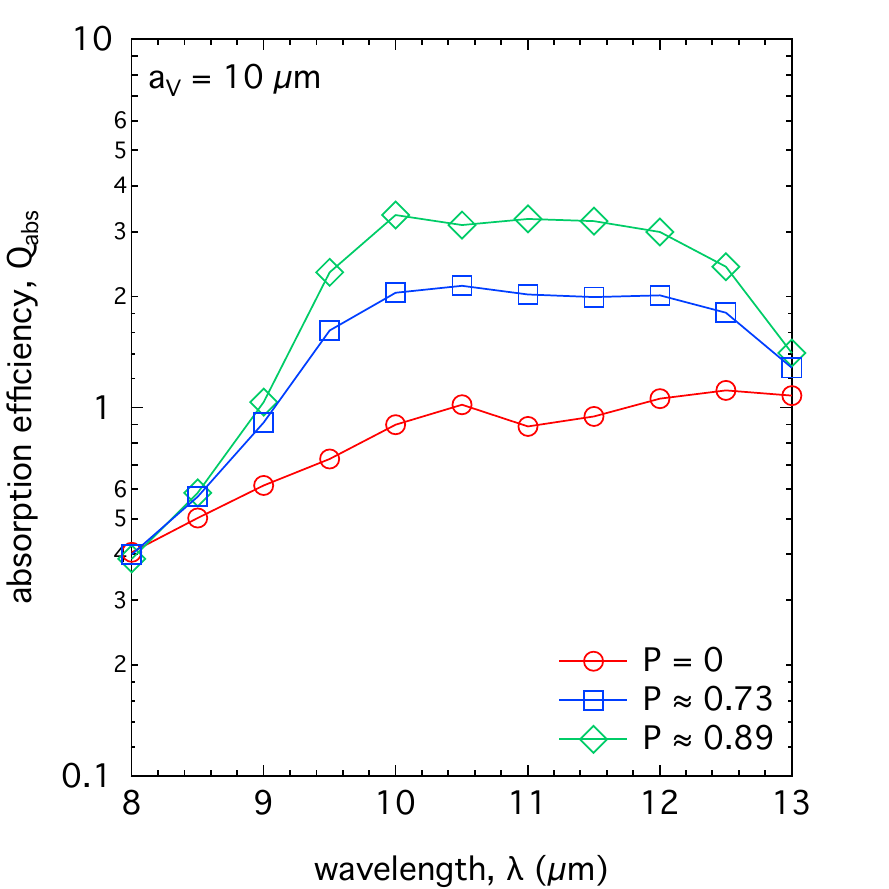}{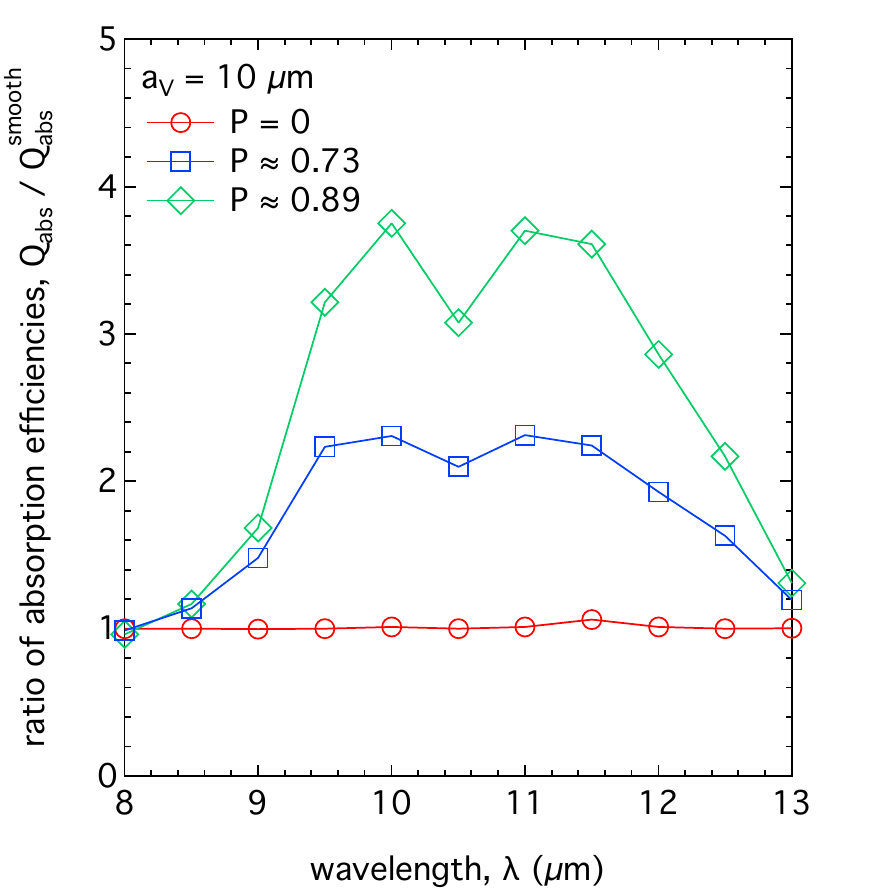}
\caption{Top: two different configurations of $N=14722$ dipoles with $d \approx 0.63~\micron$ as a porous regular tetrahedron with an inner cavity and a thin surface layer for an intermediate size of $a_\mathrm{V} = 10~\micron$.
The thickness $h$ of the surface layer is on the order of $h \approx 1.97~\micron$ (left) and $h \approx 0.66~\micron$ (right) that are translated into the porosity of $P \approx 0.73$ and $0.89$, respectively.
The two segments of the hollow tetrahedron are shown merely for the purpose of visualizing the thickness of the facets and the volume of the cavity.
Bottom: the absorption efficiency $Q_\mathrm{abs}$ (left) of tetrahedral olivine particles with rough surfaces and an inner cavity of $P \approx 0.73$ (open squares) or $0.89$ (open diamonds) and without a cavity of $P = 0$ (open circles) for an intermediate size of $a_\mathrm{V} = 10~\micron$ ($N=14722$) as a function of wavelength $\lambda$ in the range of $\lambda = 8$--$13~\micron$; the ratio of absorption efficiencies $Q_\mathrm{abs}/Q_\mathrm{abs}^\mathrm{smooth}$ (right) for tetrahedral olivine particles with rough surfaces where $Q_\mathrm{abs}^\mathrm{smooth}$ is the absorption efficiency of a tetrahedral olivine particle with $P = 0$ and smooth surfaces (i.e., $N=942714$).
\label{fig:6}}
\end{figure}
It is well known that mid-infrared spectra of highly porous silicate particles with a porosity of $P \ga 0.9$ exhibit prominent narrow emission peaks of Fr\"{o}hlich modes, even if their sizes lie in the range of $a_\mathrm{V} \ga 10~\micron$ \citep[e.g.,][]{hage-greenberg1990,okamoto-et-al1994,kimura-et-al2008,kimura-et-al2009}.
It is, however, not clear whether the Fr\"{o}hlich modes are excited by highly porous silicate aggregates only of small grains in the Rayleigh scattering regime or also of large grains with extremely flat faces.
To thoroughly investigate the effect of porosity, we first consider a single large cavity inside a regular tetrahedron with an intermediate size of $a_\mathrm{V} = 10~\micron$ and a thin surface layer.
Hollow spheres and spheroids have been occasionally used to model light scattering and thermal emission of irregularly shaped dust particles in comets and debris disks, in preference to non-porous spheres and spheroids \citep{min-et-al2005a,min-et-al2005b,milli-et-al2015}.
We should mention that a hollow structure is not unrealistic in astronomical environments, since a central cavity is a common structure of organic globules found first in the Tagish Lake carbonaceous chondrite and later in interplanetary dust particles of cometary origin and dust particles from comet 81P/Wild 2 \citep{nakamuramessenger-et-al2006,messenger-et-al2008,degregorio-et-al2009}.
Although the overall shapes of organic globules are far from a regular tetrahedron, the assumption of hollow regular tetrahedral shapes would still be useful to gain an insight into the effect of particle structures on the appearance of olivine twin peaks in thermal infrared spectra of large porous particles outside aggregates of small grains.
As depicted in the top pictures of Fig.~\ref{fig:6}, we fix the number of dipoles to $N=14722$, implying that the surface roughness is on the order of $d \approx 0.66~\micron$, and consider the surface thickness $h$ on the order of $h \approx 1.97$ (left) and $h \approx 0.66~\micron$ (right).
The two segments of the porous regular tetrahedral particles are drawn simply for the purpose of visualizing the thickness of the facets and the volume of the cavity.
By estimating the volume fraction of the cavity, we find the porosity $P$ of the regular tetrahedral particle with an inner cavity to be $P \approx 0.73$ (left) and $P \approx 0.89$ (right) for $a_\mathrm{V} = 10~\micron$ with $N=14722$.
The porosity is here defined by one minus the volume fraction of cubic dipoles that is equivalent to the total volume occupied by the dipoles in a regular tetrahedron normalized to the overall volume of the tetrahedron inclusive of the cavity.

The bottom left panel of Fig.~\ref{fig:6} compares the wavelength dependence of absorption efficiencies $Q_\mathrm{abs}$ for the porous regular tetrahedral particles of $P \approx 0.73$ (open squares) and $P \approx 0.89$ (open diamonds) with those for non-porous one of $P = 0$ (open circles), which is depicted in the upper left corner of Fig.~\ref{fig:4}.
As the porosity $P$ of regular tetrahedral particles increases, the values of $Q_\mathrm{abs}$ increase and the positions of the twin peaks are shifted towards shorter wavelengths, although the peaks are not clearly identified.
The ratios of absorption efficiencies $Q_\mathrm{abs}/Q_\mathrm{abs}^\mathrm{smooth}$ are plotted in the bottom right panel of Fig.~\ref{fig:6} where $Q_\mathrm{abs}^\mathrm{smooth}$ is again the absorption efficiency for the non-porous regular tetrahedral particle with smooth facets.
It exemplifies the formation of emission peaks in the infrared spectra of large hollow olivine particles at wavelengths of Fr\"{o}hlich modes, although they are not small grains in the Rayleigh scattering regime nor aggregates of small grains.
Accordingly, we may state that the presence of olivine twin peaks is not always an indicator of small grains and their aggregate particles in fluffy and porous configurations.
The wavelengths of the Fr\"{o}hlich modes for the hollow regular tetrahedra are marginally consistent with but slightly shorter than the peak wavelengths for the non-porous regular tetrahedron with $a_\mathrm{V} = 100~\mathrm{nm}$ shown in the bottom left panel of Fig.~\ref{fig:3}.
The difference in the peak wavelengths (i.e., the Fr\"{o}hlich frequencies) between non-porous and hollow particles most likely arises from the dependence of Fr\"{o}hlich modes on the structure of the particles, in a similar manner as the Fr\"{o}hlich modes of non-porous and hollow spheres.

While a hollow particle is distinct from an aggregate of small grains, we may regard the regular tetrahedral particles with a large cavity as an aggregate consisting of large four thin equilateral triangular prisms whose volume-equivalent radius is $6.3~\micron$.
It is worthwhile noting that the excitement of the Fr\"{o}hlich modes was indeed identified in the thermal infrared spectra of thin films measured in the laboratory \citep{baratta-et-al2000}.
Accordingly, we could attribute the presence of weak peaks in the infrared absorption spectra of hollow regular tetrahedral particles to the thinness of the constituent triangular prisms.

\subsubsection{The dispersion of open cavities}
\label{subsubsec:pores}

\begin{figure}
\plotone{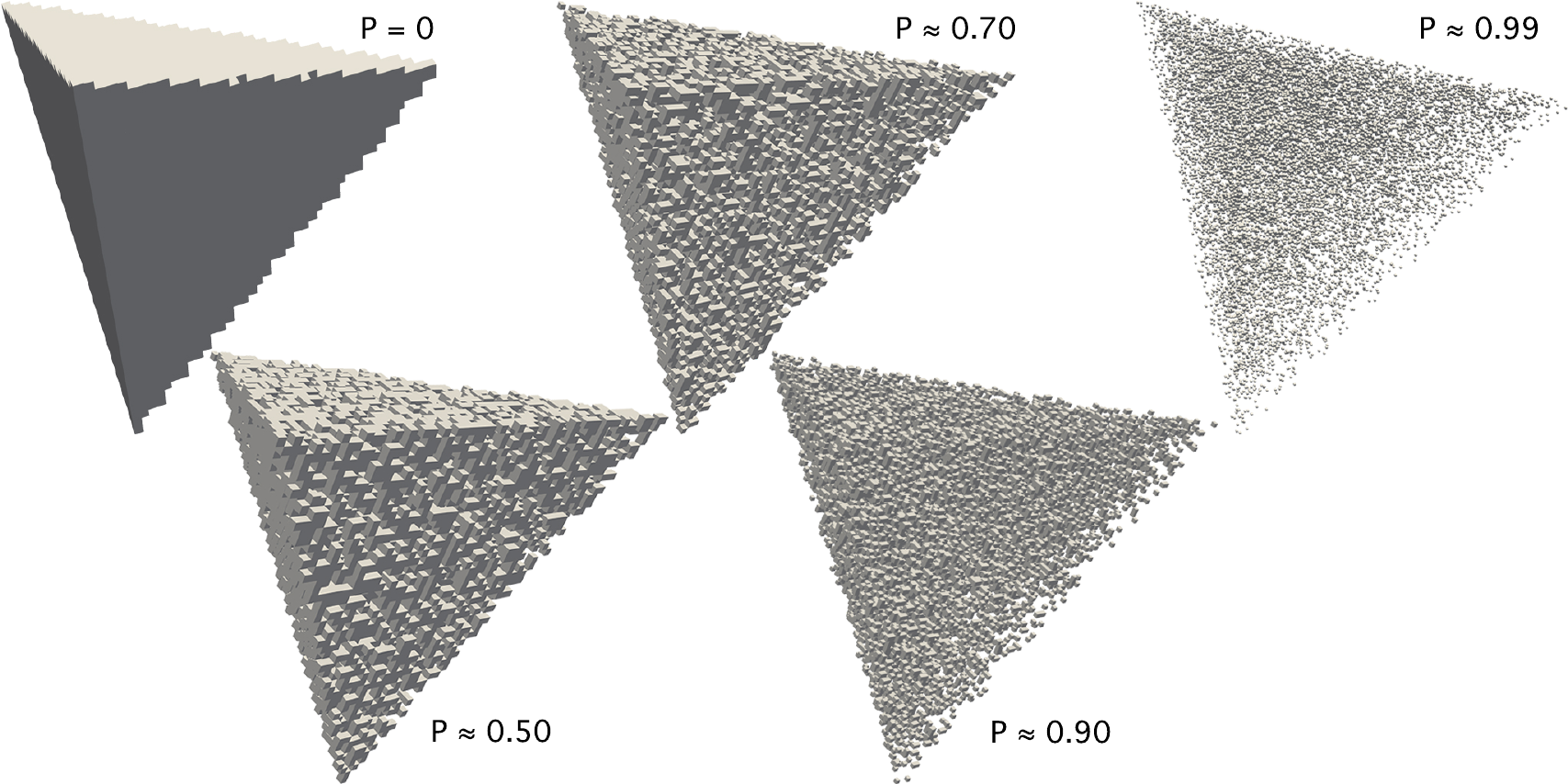}\\
\plottwo{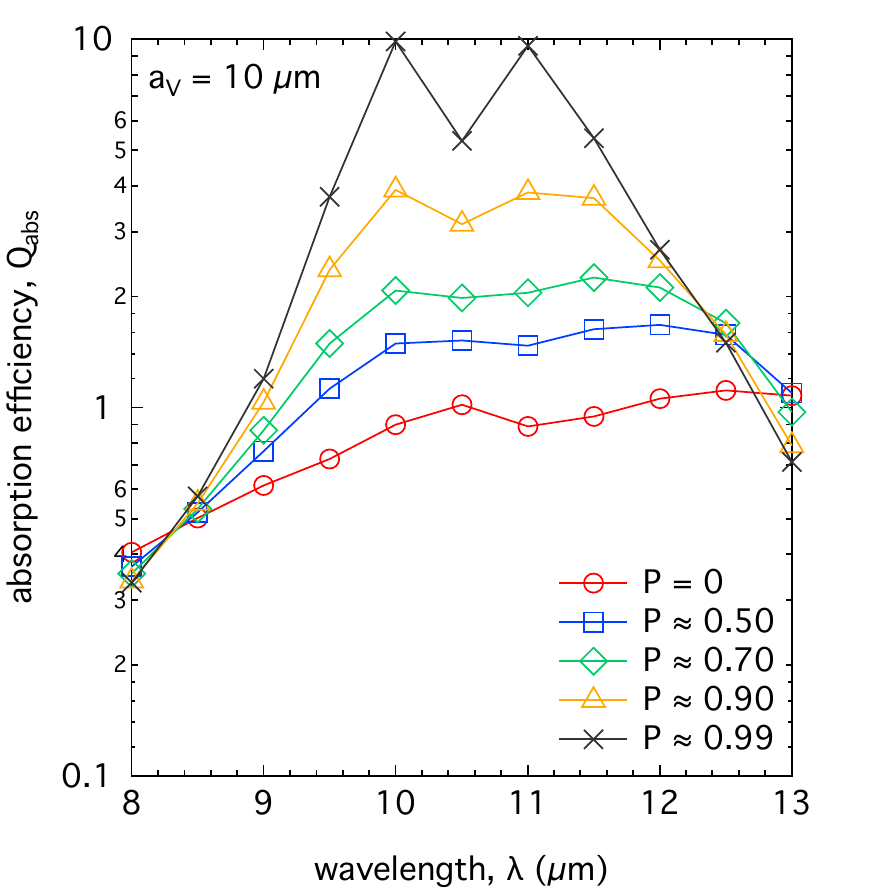}{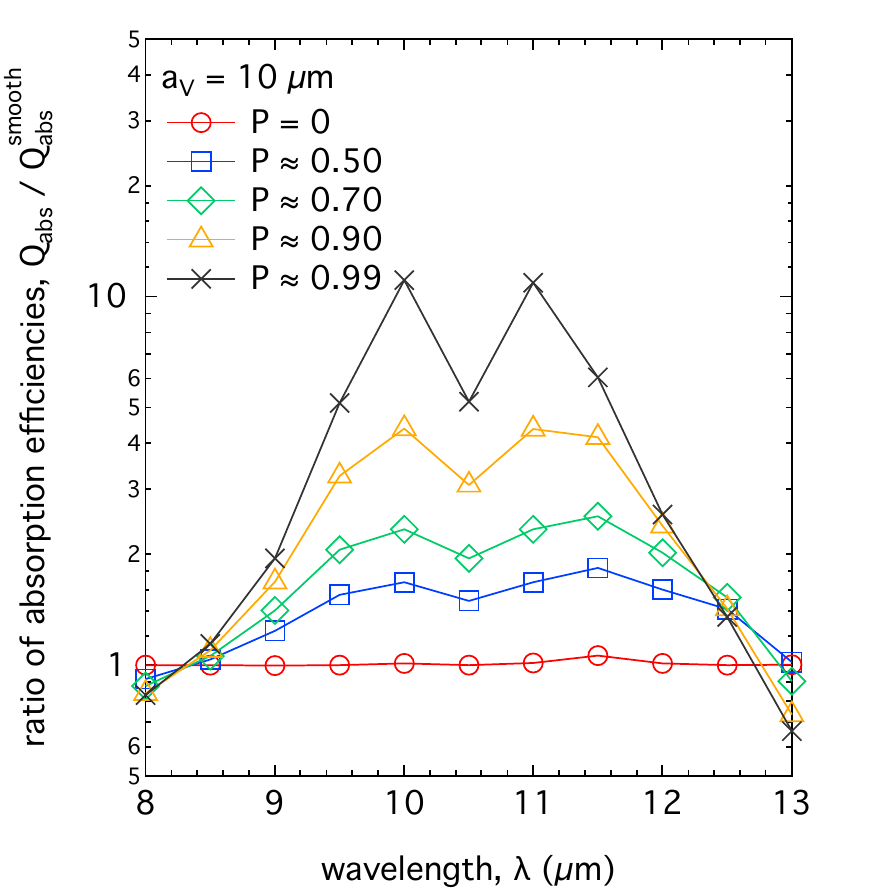}
\caption{Top: five different configurations of randomly dispersed $N=14722$ dipoles with $d \approx 0.63~\micron$ as non-porous ($P = 0 $) and porous regular tetrahedra with a porosity of $P \approx 0.50$, $ 0.70$, $0.90$, and $0.99$.
Bottom: the absorption efficiency $Q_\mathrm{abs}$ (left) of porous olivine particles with randomly dispersed dipoles and pores in a regular tetrahedron for an intermediate size of $a_\mathrm{V} = 10~\micron$ ($N=14722$) as a function of wavelength $\lambda$ in the range of $\lambda = 8$--$13~\micron$; open circles: $P=0$; open squares: $P \approx 0.50$; open diamonds: $P \approx 0.70$; open triangles: $P \approx 0.90$; crosses: $P \approx 0.99$; the ratio of absorption efficiencies $Q_\mathrm{abs}/Q_\mathrm{abs}^\mathrm{smooth}$ (right) for tetrahedral olivine particles with rough surfaces where $Q_\mathrm{abs}^\mathrm{smooth}$ is the absorption efficiency of a tetrahedral olivine particle with smooth surfaces (i.e., $N=942714$).
\label{fig:7}}
\end{figure}
Section~\ref{subsubsec:pores} is devoted to understand how the dispersal of cavities inside regular tetrahedral particles affects the broadening and shift of olivine emission peaks in the infrared spectra of the porous particles with an intermediate size of $a_\mathrm{V} = 10~\micron$ ($N=14722$ dipoles).
As depicted in the top panel of Fig.~\ref{fig:7}, we prepare porous regular tetrahedral particles consisting of $N=14722$ dipoles with $d \approx 0.66~\micron$ (Rayleigh scatterers) by randomly removing a particular number of dipoles from larger regular tetrahedra.
Since the random removal of dipoles from a regular tetrahedral particle breaks the connections between dipoles, there are a large number of dipoles that separated from the main body of the particle.
Because the presence of such ``floating'' dipoles in a regular tetrahedron is not realistic nor favorable, we later study in Sect.~\ref{subsubsec:linked} whether there is a noticeable difference in thermal infrared spectra between such a porous particle with floating dipoles and an aggregate of dipoles with a cross-linked network.

The bottom left panel of Fig.~\ref{fig:7} shows the absorption efficiencies of porous olivine particles with randomly dispersed pores and dipoles in a regular tetrahedron for an intermediate size of $a_\mathrm{V} = 10~\micron$ ($N=14722$ dipoles) as a function of wavelength $\lambda$ in the range of $\lambda = 8$--$13~\micron$.
There is a noticeable tendency that the Fr\"{o}hlich modes of Rayleigh scatterers shift the positions of the peaks toward shorter wavelengths and sharpen the peaks in the infrared spectrum of a regular tetrahedral particle, already with a porosity of $P \approx 0.5$, compared with $P = 0$.
As shown in the bottom right panel of Fig.~\ref{fig:7}, the shift of the peak wavelengths and the sharpening of the peaks are associated with excitement of Fr\"{o}hlich modes at higher porosities.
These tendencies become more remarkable, when the porosity is elevated to $P \approx 0.70$, $0.90$, and $0.99$, as expected for porous aggregates of small constituent grains as far as minute grains in the Rayleigh scattering regime are concerned.
However, the peaks for the intermediate-sized highly porous ($P \ga 0.9$) regular tetrahedron with randomly dispersed pores appear at shorter wavelengths, compared to the tiny non-porous tetrahedron of $a_\mathrm{V} \le 100~\mathrm{nm}$.
Here, we should emphasize that the tiny non-porous particle of $a_\mathrm{V} = 100~\mathrm{nm}$ has the overall shape of tetrahedron, while each dipole in the highly porous particle does not.
We may, therefore, attribute the difference in the wavelengths of Fr\"{o}hlich modes to the effect of grain shape on the absorption spectra of an ensemble of grains or an aggregate of grains.
It is also worthwhile noting that the Fr\"{o}hlich modes of the intermediate-sized highly porous ($P \ga 0.9$) regular tetrahedron with randomly dispersed pores are excited at shorter wavelengths, compared to the non-porous regular tetrahedron with rigged surfaces (cf. the bottom right panel of Fig.~\ref{fig:4}).
The dipoles that represent surface bumps of the non-porous particles are essentially identical to the dipoles that constitute an ensemble of dipoles in the porous particles, while the space between the dipoles is to a large extent occupied by other dipoles in the former, but pores in the latter.
Accordingly, the infrared spectra of small particles and their ensembles in fluffy and porous configurations depend on the surroundings in which they excite the Fr\"{o}hlich modes.

\subsubsection{A cross-linked network}
\label{subsubsec:linked}

\begin{figure}
\plottwo{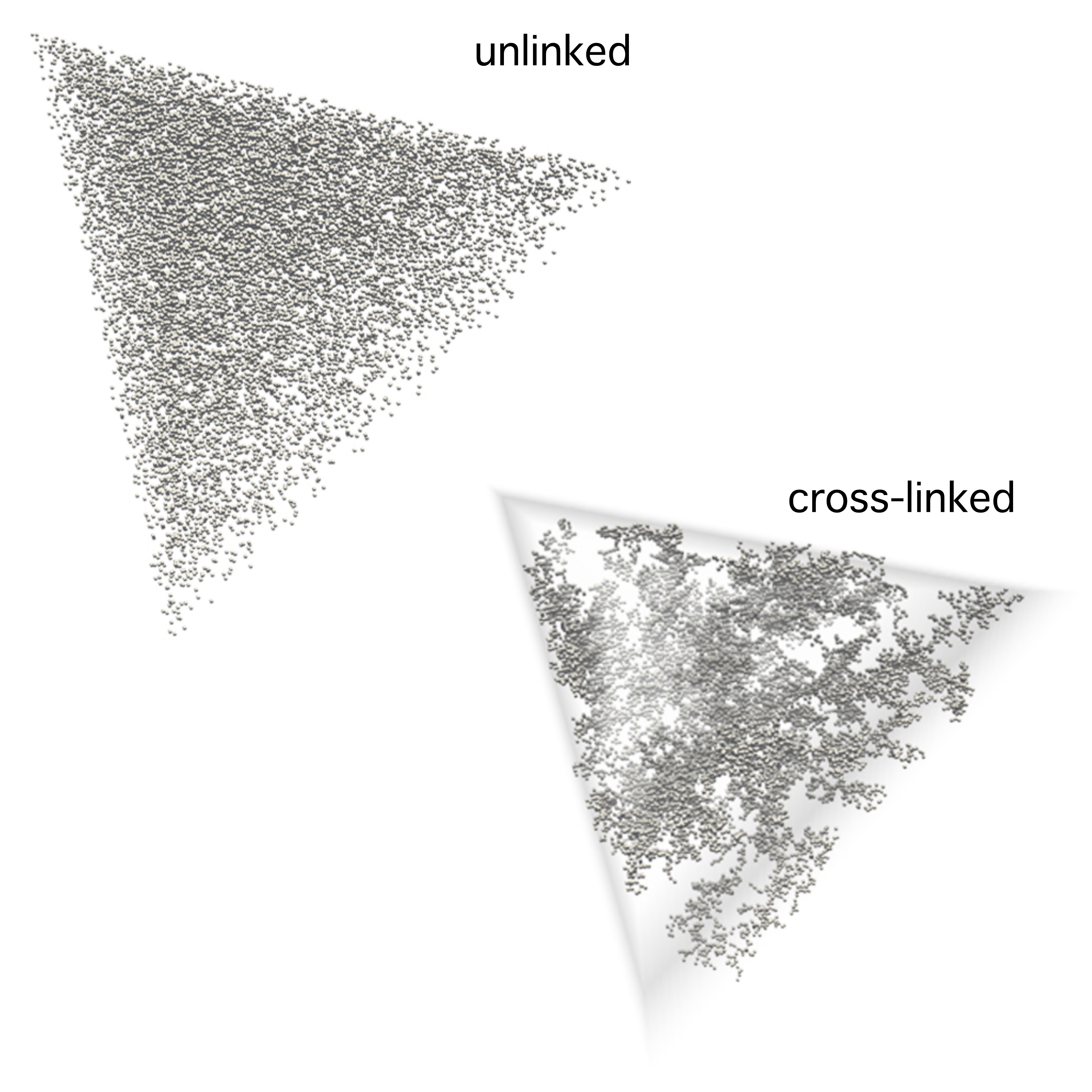}{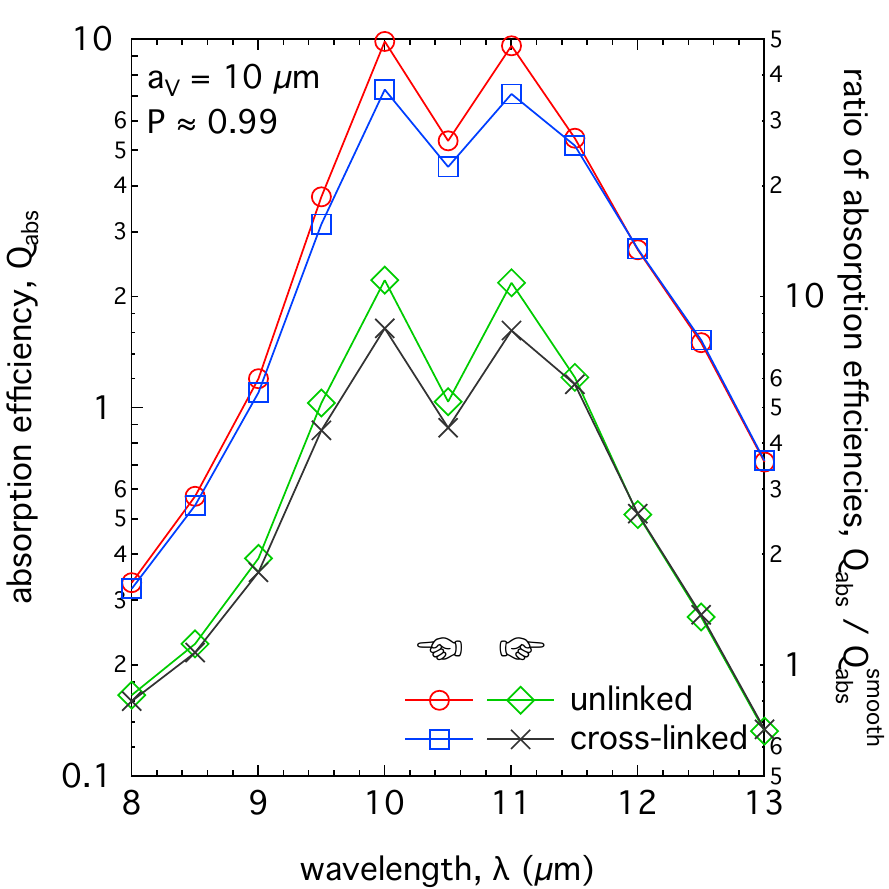}
\caption{Left: two different configurations of $N=14722$ dipoles in a regular tetrahedral shape with a volume-equivalent sphere radius of $a_\mathrm{V} = 10~\micron$, a lattice edge length of $d \approx 0.63~\micron$, and a porosity of $P \approx 0.99$. 
The dipole locations are either randomly distributed in a regular tetrahedral shape (unlinked) or determined initially by diffusion-limited aggregation (DLA) and later chopped off into a regular tetrahedral shape (cross-linked).
The snippet of the DLA particle is overlaid with an opaque regular tetrahedral shape to guide eyes and cavities inside the regular tetrahedron defines the porosity $P$ of the snippety particle.
Right: the absorption efficiency $Q_\mathrm{abs}$ of regular tetrahedral olivine particles with a volume-equivalent sphere radius of $a_\mathrm{V} = 10~\micron$, a lattice edge length of $d \approx 0.63~\micron$, and a porosity of $P \approx 0.99$ (left axis) as a function of wavelength $\lambda$ in the range of $\lambda = 8$--$13~\micron$;
open circles: an unlinked network of dipoles; open squares: a cross-linked network of dipoles.
Also plotted (right axis) is the wavelength dependence of absorption efficiency normalized to the absorption efficiency $Q_\mathrm{abs}^\mathrm{smooth}$ of a non-porous olivine particle with a regular tetrahedral shape and smooth surfaces (i.e., $N=942714$): open diamonds: an unlinked network of dipoles ; crosses: a cross-linked network of dipoles.
\label{fig:8}}
\end{figure}

At last, we present our numerical results of absorption efficiencies for highly porous particles with a cross-linked network of Rayleigh scatterers (dipoles) that are confined in a regular tetrahedral shape.
The configuration of dipoles is arranged firstly with diffusion-limited aggregation (DLA), secondly with the cutoff of dipoles that are located outside a regular tetrahedral shape, and finally with the removal of dipoles that are detached from the main body as a result of the cutoff.
The left panel of Fig.~\ref{fig:8} shows the configurations of $N=14722$ dipoles in a regular tetrahedral shape with a radius of $a_\mathrm{V} = 10~\micron$, a lattice edge length of $d \approx 0.63~\micron$, and a porosity of $P \approx 0.99$ where the cluster of unlinked floating dipoles is identical to the most porous one used in Sect.~\ref{subsubsec:pores}.
The snippet of the DLA particle is overlaid with an opaque regular tetrahedron to guide eyes and cavities inside the tetrahedron, which is configured by $N = 1396696$ dipoles with $d \approx 0.63~\micron$, defines the porosity $P$ of the snippety particle.

The right panel of Fig.~\ref{fig:8} depicts the effect of cross-links on the thermal infrared spectra of highly porous particles arranged in a regular tetrahedral shape.
We find that results with an ensemble of floating dipoles (open circles) do not significantly differ from an aggregate of dipoles with a cross-linked network (open squares) as far as absorption efficiencies for highly porous particles are concerned.
Slight distinctions in absorption efficiencies may be noticeable near the twin peaks, but the wavelengths of the peaks do not seem to be affected by the presence of links between dipoles.
Also plotted in the right panel of Fig.~\ref{fig:8} are the values for floating dipoles (open diamonds) and cross-linked dipoles (crosses) normalized to the absorption efficiency $Q_\mathrm{abs}^\mathrm{smooth}$ of a non-porous olivine particle with a regular tetrahedral shape and smooth surfaces (i.e., $N=942714$).
By the same token, the results of normalized absorption efficiencies for an ensemble of floating dipoles and an aggregate of dipoles with a cross-linked network resemble each other closely.
Therefore, cross-links between dipoles do not play a vital role in the determination of Fr\"{o}hlich frequencies, at the very least, for highly porous ($P \approx 0.99$) particles as shown in the left panel of Fig.~\ref{fig:8}.

In contrast to the peak wavelengths, the strengths of the twin peaks are slightly higher for the unlinked network of dipoles than the cross-linked network of dipoles.
We may attribute the slight enhancement in the peaks to the difference in the surroundings of the dipoles, by considering that the dipole in the former are typically embedded in vacuum and those in the latter have at least one neighboring dipole.
It is worthwhile noting that such a cloud of dipoles has been used to justify the applicability of effective medium theories to calculate absorption efficiencies of highly porous dust aggregates \citep{hage-greenberg1990,okamoto-et-al1994}.
We should, therefore, mention that the use of effective medium theories to the computation of absorption efficiencies slightly overestimates the strength of absorptioin peaks in thermal infrared spectra of dust aggregates.

\section{Concluding remarks}
\label{sec:remarks}

Our theoretical argument and numerical verification secure sufficient evidence to reaffirm that the presence of strong absorption peaks in the infrared spectra of non-porous olivine particles is a diagnostic character of small ($a_\mathrm{V} \ll 10~\micron$) particle size. 
To be exact, the spectral variation in the absorption efficiency of non-porous olivine particles is not completely featureless even in the geometrical optics limit ($a_\mathrm{V} \ga 1~\mathrm{mm}$).
As a result, there is room for making faint features clearly visible in laboratory experiments, when the infrared spectra of dust particles with $Q_\mathrm{abs} \simeq 1$ is magnified and the noise level is dampened simultaneously.
Here, we should emphasize that faint features in the infrared spectra of large non-porous particles with the geometrical optics regime are troughs, instead of peaks, as depicted in the bottom left panel of Fig.~\ref{fig:3}.
This agrees with mid-infrared spectra of thermal emission from a slab of silicate samples previously measured in the laboratory where troughs appear in the emission spectra as well as the spectra of $1-R$ \citep{salisbury-et-al1994,lane-et-al2011,bramble-et-al2019,bramble-et-al2021}.
Any attempts to identify faint troughs of large non-porous silicate particles in remote-sensing observational data would, however, suffer from low signal-to-noise ratios, because a predominance of large particles is likely associated with very weak signals due to their low spatial densities.
Nonetheless, silicate emission troughs could be identified in the thermal infrared spectra of cometary nuclei, a successful example of which is comet P/2016 BA14 (PANSTARRS), when they are measured under ideal circumstances of observation with a large telescope at very small geocentric distances \citep{ootsubo-et-al2021}.
All in all, light-scattering theory, numerical simulation, laboratory experiments, and remote-sensing observations point to the outcome of surface modes that strong silicate emission peaks of particles turn into weak troughs as the size of the particles increases.
Therefore, there is no evidence to refute the consensus that the appearance of silicate emission peaks in the infrared spectra of cometary comae is associated with small sizes for non-porous particles.

If dust particles are large ($a_\mathrm{V} \ga 10~\micron$) aggregates consisting of small grains, then the presence of $10~\micron$ silicate peaks must be attributed to a minute size of constituent grains and a high porosity of the aggregates.
Weak emission peaks of Mg-rich olivine appeared in the infrared spectrum of dust aggregates in the coma of comet 67P/Churyumov-Gerasimenko, while the size of the constituent grains and the porosity of the aggregates were measured in situ by various instruments onboard Rosetta \citep{wooden-et-al2017}.
Detailed analysis of topographic images taken by MIDAS revealed that the size of constituent grains in the aggregates is on the order of sub-micrometers and the structure of the aggregates is described as a fractal \citep{bentley-et-al2016,mannel-et-al2016}.
It turned out that the porosity $P \simeq 0.85$ of fractal particles formed under the ballistic particle-cluster aggregation (BPCA) process is consistent with in-situ data of MIDAS, OSIRIS, GIADA, and COSIMA for dust in the coma and on the surface of the comet nucleus \citep{fornasier-et-al2015,kimura-et-al2020a}.
Moreover, the porosity of $P = 0.87$, comparable with the one expected for the BPCA fractal particles, well accounts for the tensile strengths of dust particles and overhangs derived from optical images taken by COSIMA and OSIRIS \citep{kimura-et-al2020b}.
All the in-situ dust measurements of the Rosetta mission are in good harmony with a model for the formation of comets that predicts the hierarchical structure of dust aggregates due to the so-called rainout growth under the BPCA process subsequent to the ballistic cluster-cluster aggregation (BCCA) process \citep{weidenschilling1997}.
Apparently, the Rosetta mission provided convincing evidence that thermal emission from large aggregates having a radius of $a_\mathrm{V} \ga 10~\micron$ and a porosity of $P \approx 0.9$ and consisting of submicron grains could still produce silicate peaks in the infrared spectra.

\begin{acknowledgments}
We would like to thank Takafumi Ootsubo for useful discussion on the detectability of silicate emission features in the infrared spectra of cometary comae by ground-based and space-borne telescopic observations.
We are indebted to Bruce T. Draine and Piotr J. Flatau for their generosities of making the Fortran 90 code ``DDSCAT'' publicly available and their efforts to continuously improve the code.
H.K is supported by the Grants-in-Aid for Scientific Research (KAKENHI JP21H00050) of Japan Society for the Promotion of Science (JSPS) and J.M. by the European Union’s Horizon 2020 research and innovation programme under grant agreement No 757390 CAstRA.
\end{acknowledgments}




\begin{thebibliography}{}

\bibitem[Anderson(2003)]{anderson2003}
Anderson, M. S.\ 2003. Appl. Phys. Lett.. 83, \href{https://doi.org/10.1063/1.1615317}{2964--2966}

\bibitem[Aronson \& Emslie(1975)]{aronson-emslie1975}
Aronson, J. R., \& Emslie, A. G.\ 1975. \jgr, 80, \href{https://doi.org/10.1029/JB080i035p04925}{4925--4931}

\bibitem[Baratta et al.(2000)]{baratta-et-al2000}
Baratta, G. A., Palumbo, M. E., \& Strazzulla, G.\ 2000. \aap, 357, 1045--1050.

\bibitem[Beck et al.(2021)]{beck-et-al2021}
Beck, P., Schmitt, B., Potin, S., Pommerol, A., \& Brissaud, O.\ 2021. Icarus, 354, \href{https://doi.org/10.1016/j.icarus.2020.114066}{114066}

\bibitem[Bentley et al.(2016)]{bentley-et-al2016}
Bentley, M. S., Schmied, R., Mannel, T., Torkar, K., Jeszenszky, H., Romstedt, J., et al.\ 2016. Nature, \href{https://doi.org/10.1038/nature19091}{537, 73--75}

\bibitem[Bertini et al.(2009)]{bertini-et-al2009}
Bertini, I., Gutierrez, P. J., \& Sabolo, W.\ 2009. \aap, 504, \href{https://doi.org/10.1051/0004-6361/200912248}{625--633}

\bibitem[Bohren \& Huffman(1983)]{bohren-huffman1983}
Bohren, C.~F., \& Huffman, D.~R.\ 1983. Absorption and Scattering of Light by Small Particles (New York: Wiley-Interscience)

\bibitem[Bramble et al.(2019)]{bramble-et-al2019}
Bramble, M. S., Yang, Y., Patterson III, W. R., Milliken, R. E., Mustard, J. F., \& Donaldson Hanna, K. L.\ 2019. Rev. Sci. Instrum., 90, \href{https://doi.org/10.1063/1.5096363}{093101}

\bibitem[Bramble et al.(2021)]{bramble-et-al2021}
Bramble, M. S., Milliken, R. E., \& Patterson III, W. R.\ 2021. Icarus, 369, \href{https://doi.org/10.1016/j.icarus.2021.114561}{114561}

\bibitem[Bregman et al.(1987)]{bregman-et-al1987}
Bregman, J. D., Campins, H., Witteborn, F. C., Wooden, D. H., Rank, D. M., Allamandola, L. J., et al.\ 1987. \aap, 187, 616--620.

\bibitem[Chang et al.(2005)]{chang-et-al2005}
Chang, P. C. Y., Walker, J. G., \& Hopcraft, K. I.\ 2005. \jqsrt, \href{https://doi.org/10.1016/j.jqsrt.2005.01.001}{96, 327--341}

\bibitem[Chornaya et al.(2020)]{chornaya-et-al2020}
Chornaya, E., Zakharenko, A. M., Zubko, E., Kuchmizhak, A., Golokhvast, K. S., \& Videen, G.\ 2020. Icarus, \href{https://doi.org/10.1016/j.icarus.2020.113907}{350, 113907}

\bibitem[De Gregorio et al.(2009)]{degregorio-et-al2009}
De Gregorio, B. T., Stroud, R. M., Nittler, L. R., Cody, G. D., \& Kilcoyne, A. L. D.\ 2009. 40th Annual Lunar and Planetary Science Conference, p. 1130.

\bibitem[Draine(1989)]{draine1989}
Draine, B., 1989\ in Interstellar Dust, ed. L. J. Allamandola, \& A.G.G.M., Tielens (Dordrecht: Kluwer Academic Publishers), pp. 313--327

\bibitem[Draine \& Flatau(1994)]{draine-flatau1994}
Draine, B. T., \& Flatau, P. J.\ 1994. J. Opt. Soc. Am. A, \href{https://doi.org/10.1364/JOSAA.11.001491}{11, 1491--1499}

\bibitem[Draine \& Flatau(2008)]{draine-flatau2008}
Draine, B. T., \& Flatau, P. J.\ 2008. J. Opt. Soc. Am. A, 25, \href{https://doi.org/10.1364/JOSAA.25.002693}{2693--2703}

\bibitem[Fornasier et al.(2015)]{fornasier-et-al2015}
Fornasier, S., Hasselmann, P. H., Barucci, M. A., Feller, C., Besse, S., Leyrat, C., et al.\ 2015, \aap, \href{https://doi.org/10.1051/0004-6361/201525901}{583, A30}

\bibitem[Fr\"{o}hlich(1949)]{froehlich1949}
Fr\"{o}hlich, H., 1949. Theory of Dielectrics (London: Oxford University Press)

\bibitem[Fuchs(1974)]{fuchs1974}
Fuchs, R.\ 1974. Phys. Lett. A, 48, \href{https://doi.org/10.1016/0375-9601(74)90463-0}{353--354}

\bibitem[Gilra(1972)]{gilra1972}
Gilra, D. P.\ 1972. In Scientific Results from the Orbiting Astronomical Observatory (OAO-2) (SP-310), ed. A. D. Code (Washington: NASA), pp. 295--319.

\bibitem[Greenberg \& Hage(1990)]{greenberg-hage1990}
Greenberg, J. M., \& Hage, J. I.\ 1990. \apj, \href{https://doi.org/10.1086/169191}{361, 260--274}

\bibitem[Gutkowicz-Krusin \& Draine(2004)]{gutkowiczkrusin-draine2004}
Gutkowicz-Krusin, D., \& Draine, B. T.\ 2004. \href{https://arxiv.org/abs/astro-ph/0403082}{eprint arXiv:astro-ph/0403082}

\bibitem[Hage \& Greenberg(1990)]{hage-greenberg1990}
Hage, J. I., \& Greenberg, J. M.\ 1990. \apj, 361, \href{https://doi.org/10.1086/169190}{251--259}

\bibitem[Hanner \& Bradley(2004)]{hanner-bradley2004}
Hanner, M. S., \& Bradley, J. P.\ 2004. In Comets II, ed. M. C. Festou, H. U. Keller, \& H. A. Weaver (Tucson: The University of Arizona Press), pp. 555--564.

\bibitem[Hanner et al.(1981)]{hanner-et-al1981}
Hanner, M. S., Giese, R. H., Weiss, K., \& Zerull, R.\ 1981. \aap, 104, 42--46.

\bibitem[Hanner et al.(1987)]{hanner-et-al1987}
Hanner, M. S., Tokunaga, A. T., Golisch, W. F., Griep, D. M., \& Kaminski, C. D.\ 1987. \aap, 187, 653--660

\bibitem[Hanner et al.(1997)]{hanner-et-al1997}
Hanner, M. S., Gehrz, R. D., Harker, D. E., Hayward, T. L., Lynch, D. K., Mason, C. C., et al.\ 1997. Earth, Moon, and Planets, 79, \href{https://doi.org/10.1023/A:1006201820477}{247--264}

\bibitem[Hapke(2012)]{hapke2012}
Hapke, B.\ 2012. Theory of Reflectance and Emittance Spectroscopy, 2nd Ed. (Cambridge: Cambridge University Press)

\bibitem[Harker et al.(2002)]{harker-et-al2002}
Harker, D. E., Wooden, D. H., Woodward, C. E., \& Lisse, C. M.\ 2002. \apj, 580, \href{https://doi.org/10.1086/343091}{579--597}

\bibitem[Hayashi(1984)]{hayashi1984}
Hayashi, S.\ 1984. Jpn. J. Appl. Phys., 23, \href{https://doi.org/10.1143/JJAP.23.665}{665--676}

\bibitem[Huffman(1975)]{huffman1975}
Huffman, D. R.\ 1975. Astrophys. Space Sci., 34, \href{https://doi.org/10.1007/BF00646757}{175--184}

\bibitem[Huffman(1977)]{huffman1977}
Huffman, D. R.\ 1977. Adv. Phys., 26, \href{https://doi.org/10.1080/00018737700101373}{129--230}

\bibitem[Hunt(1976)]{hunt1976}
Hunt, G. R.\ 1976. J. Phys. Chem., \href{https://doi.org/10.1021/j100552a015}{80, 1195--1198}

\bibitem[Hunt \& Logan(1972)]{hunt-logan1972}
Hunt, G. R., \& Logan, L. M.\ 1972. Appl. Opt. \href{https://doi.org/10.1364/AO.11.000142}{11, 142--147}

\bibitem[Ishizuka et al.(2018)]{ishizuka-et-al2018}
Ishizuka, S., Kimura, Y., Sakon, I., Kimura, H., Yamazaki, T., Takeuchi, S., \& Inatomi, Y.\ 2018. Nat. Commun., 9, \href{https://doi.org/10.1038/s41467-018-06359-y}{3820}

\bibitem[Jackson(1998)]{jackson1998}
Jackson, J. D.\ 1998. Classical Electrodynamics, 3rd Ed. (Hoboken, NJ: John Wiley \& Sons, Inc.)

\bibitem[Jurewicz et al.(2003)]{jurewicz-et-al2003}
Jurewicz, A., Orofino, V., Marra, A. C., \& Blanco, A.\ 2003. \aap, 410, \href{https://doi.org/10.1051/0004-6361:20031317}{1055--1062}

\bibitem[Kimura(2014)]{kimura2014}
Kimura, H.\ 2014. Icarus, 232, \href{https://doi.org/10.1016/j.icarus.2014.01.009}{133--140}

\bibitem[Kimura et al.(2003)]{kimura-et-al2003}
Kimura, H., Kolokolova, L., \& Mann, I.\ 2003. \aap, \href{https://doi.org/10.1051/0004-6361:20030967}{407, L5--L8}

\bibitem[Kimura et al.(2006)]{kimura-et-al2006}
Kimura, H., Kolokolova, L., \& Mann, I.\ 2006. \aap, \href{https://doi.org/10.1051/0004-6361:20041783}{449, 1243--1254}

\bibitem[Kimura et al.(2008)]{kimura-et-al2008}
Kimura, H., Chigai, T., \& Yamamoto, T.\ 2008. \aap, \href{https://doi.org/10.1051/0004-6361:20078778}{482, 305--307}

\bibitem[Kimura et al.(2009)]{kimura-et-al2009}
Kimura, H., Chigai, T., \& Yamamoto, T.\ 2009. \apj, 690, \href{https://doi.org/10.1088/0004-637X/690/2/1590}{1590--1596}

\bibitem[Kimura et al.(2020a)]{kimura-et-al2020a}
Kimura, H., Hilchenbach, M., Merouane, S., Paquette, J., \& Stenzel, O.\ 2020a. \planss, \href{https://doi.org/10.1016/j.pss.2019.104825}{181, 104825}

\bibitem[Kimura et al.(2020b)]{kimura-et-al2020b}
Kimura, H., Wada, K., Yoshida, F., Hong, P. K., Senshu, H., Arai, T., et al.\ 2020b. \mnras, \href{https://doi.org/10.1093/mnras/staa1641}{496, 1667--1682}

\bibitem[Knacke(1968)]{knacke1968}
Knacke, R. F.\ 1968. Nature, 217, \href{https://doi.org/10.1038/217044a0}{44--45}

\bibitem[Knacke et al.(1993)]{knacke-et-al1993}
Knacke, R. F., Fajardo-Acosta, S. B., Telesco, C. M., Hackwell, J. A., Lynch, D. K., \& Russell, R. W.\ 1993. \apj, 418, \href{https://doi.org/10.1086/173405}{440--450}

\bibitem[Koike et al.(2003)]{koike-et-al2003}
Koike, C., Chihara, H., Tsuchiyama, A., Suto, H., Sogawa, H., \& Okuda, H.\ 2003. \aap, 399, \href{https://doi.org/0004-6361:20021831}{1101--1107}

\bibitem[Koike et al.(2010)]{koike-et-al2010}
Koike, C., Imai, Y., Chihara, H., Suto, H., Murata, K., Tsuchiyama, A., et al.\ 2010. \apj, 709, \href{https://doi.org/10.1088/0004-637X/709/2/983}{983--992}

\bibitem[Kolokolova et al.(2007)]{kolokolova-et-al2007}
Kolokolova, L., Kimura, H., Kiselev, N., \& Rosenbush, V.\ 2007. \aap, \href{https://doi.org/10.1051/0004-6361:20065069}{463, 1189--1196}

\bibitem[Krishna Swamy \& Donn(1979)]{krishnaswamy-donn1979}
Krishna Swamy, K. S., \& Donn, B.\ 1979. \aj, \href{https://doi.org/10.1086/112469}{84, 692--697}

\bibitem[Lane et al.(2011)]{lane-et-al2011}
Lane, M. D., Glotch, T. D., Dyar, M. D., Pieters, C. M., Klima, R., et al.\ 2011. \jgr, \href{https://doi.org/10.1029/2010JE003588}{116, E08010}

\bibitem[Lindqvist et al.(2018)]{lindqvist-et-al2018}
Lindqvist, H., Martikainen, J., R\"{a}bin\"{a}, J., Penttil\"{a}, A., \& Muinonen, K. 2018. \jqsrt, \href{https://doi.org/10.1016/j.jqsrt.2018.06.005}{217, 329--337}

\bibitem[Mannel et al.(2016)]{mannel-et-al2016}
Mannel, T., Bentley, M. S., Schmied, R., et al.\ 2016. \mnras, 462, \href{https://doi.org/10.1093/mnras/stw2898}{S304--S311}

\bibitem[Messenger et al.(2008)]{messenger-et-al2008}
Messenger, S., Nakamura-Messenger, K., \& Keller, L. P.\ 2008. 39th Annual Lunar and Planetary Science Conference, p. 2391.

\bibitem[Milli et al.(2015)]{milli-et-al2015}
Milli, J., Mawet, D., Pinte, C., Lagrange, A.-M., Mouillet, D., Girard, J. H., et al.\ 2015. \aap, 577, \href{https://doi.org/10.1051/0004-6361/201423950}{A57}

\bibitem[Min et al.(2003)]{min-et-al2003}
Min, M., Hovenier, J. W., \& de Koter, A.\ 2003. \jqsrt, 79--80, \href{https://doi.org/10.1016/S0022-4073(02)00330-8}{939--951}

\bibitem[Min et al.(2005a)]{min-et-al2005a}
Min, M., Hovenier, J. W., de Koter, A., Waters, L. B. F. M., \& Dominik, C.\ 2005a. Icarus, 179, \href{https://doi.org/10.1016/j.icarus.2005.05.015}{158--173}

\bibitem[Min et al.(2005b)]{min-et-al2005b}
Min, M., Hovenier, J. W., \& de Koter, A.\ 2005b. \aap, 432, \href{https://doi.org/10.1051/0004-6361:20041920}{909--920}

\bibitem[Mukai \& Koike(1990)]{mukai-koike1990}
Mukai, T., \& Koike, C.\ 1990. Icarus, \href{https://doi.org/10.1016/0019-1035(90)90027-7}{87, 180--187}

\bibitem[Mukai et al.(1992)]{mukai-et-al1992}
Mukai, T., Ishimoto, H., Kozasa, T., Blum, J., \& Greenberg, J. M.\ 1992. \aap, 262, 315--320.

\bibitem[Nakamura-Messenger et al.(2006)]{nakamuramessenger-et-al2006}
Nakamura-Messenger, K., Messenger, S., Keller, L. P., Clemett, S. J., \& Zolensky, M. E.\ 2006. Science, 314, \href{https://doi.org/10.1126/science.1132175}{1439--1442}

\bibitem[Nussenzveig \& Wiscombe(1980)]{nussenzveig-wiscombe1980}
Nussenzveig, H. M., \& Wiscombe, W. J.\ 1980. \prl, 45, \href{https://doi.org/10.1103/PhysRevLett.45.1490}{1490--1495}

\bibitem[Okamoto et al.(1994)]{okamoto-et-al1994}
Okamoto, H., Mukai, T., \& Kozasa, T.\ 1994. \planss, \href{https://doi.org/10.1016/0032-0633(94)90041-8}{42, 643--649}

\bibitem[Ootsubo et al.(2007)]{ootsubo-et-al2007}
Ootsubo, T., Watanabe, J.-I., Kawakita, H., Honda, M., \& Furusho, R.\ 2007. \planss, 55, \href{https://doi.org/10.1016/j.pss.2006.11.012}{1044--1049}

\bibitem[Ootsubo et al.(2021)]{ootsubo-et-al2021}
Ootsubo, T., Kawakita, H., \& Shinnaka, Y.\ 2021. Icarus, 363, \href{https://doi.org/10.1016/j.icarus.2021.114425}{114425}

\bibitem[Peake(1959)]{peake1959}
Peake, W. H.\ 1959. IRE Trans. Antennas Propag., \href{https://doi.org/10.1109/TAP.1959.1144736}{7, 324--329}

\bibitem[Pitman et al.(2013)]{pitman-et-al2013}
Pitman, K. M., Hofmeister, A. M., \& Speck, A. K.\ 2013. Earth Planets Space, 65, \href{https://doi.org/10.5047/eps.2012.05.009}{129--138}

\bibitem[Rea \& Welch(1963)]{rea-welch1963}
Rea, D. G., \& Welch, W. J.\ 1963. \ssr, \href{https://doi.org/10.1007/BF00172384}{2, 558--617}

\bibitem[Reach et al.(2003)]{reach-et-al2003}
Reach, W. T., Morris, P., Boulanger, F., \& Okumura, K.\ 2003. Icarus, 164, \href{https://doi.org/10.1016/S0019-1035(03)00133-7}{384--403}

\bibitem[Rose(1979)]{rose1979}
Rose, L. A.\ 1979. Astrophys. Space Sci., \href{https://doi.org/10.1007/BF00643489}{65, 47--67}

\bibitem[Ruppin(1997)]{ruppin1997}
Ruppin, R.\ 1997. Phys. Lett. A, \href{https://doi.org/10.1016/S0375-9601(97)00735-4}{237, 99--102}

\bibitem[Ruppin \& Englman(1970)]{ruppin-englman1970}
Ruppin, R., \& Englman, R.\ 1970. Rep. Prog. Phys., 33, \href{https://doi.org/10.1088/0034-4885/33/1/304}{149--196}

\bibitem[Salisbury \& Wald(1992)]{salisbury-wald1992}
Salisbury, J. W., \& Wald, A.\ 1992. Icarus, 96, \href{https://doi.org/10.1016/0019-1035(92)90009-V}{121--128}

\bibitem[Salisbury et al.(1994)]{salisbury-et-al1994}
Salisbury, J. W., Wald, A., \& D'Aria, D. M.\ 1994. \jgr, 99, \href{https://doi.org/10.1029/93JB03600}{11897--11911}

\bibitem[Schlick(1994)]{schlick1994}
Schlick, C.\ 1994. Comput. Graph. Forum, \href{https://doi.org/10.1111/1467-8659.1330233}{13, 233--246}

\bibitem[Sogawa et al.(2006)]{sogawa-et-al2006}
Sogawa, H., Koike, C., Chihara, H., Suto, H., Tachibana, S., Tsuchiyama, A., \& Kozasa, T.\ 2006. \aap, \href{https://doi.org/10.1051/0004-6361:20041538}{451, 357--361}

\bibitem[Steyer et al.(1974)]{steyer-et-al1974}
Steyer, T. R., Day, K. L., \& Huffman, D. R.\ 1974. \ao, 13, \href{https://doi.org/10.1364/AO.13.001586}{1586--1590}

\bibitem[Weidenschilling(1997)]{weidenschilling1997}
Weidenschilling, S. J.\ 1997. Icarus, \href{https://doi.org/10.1006/icar.1997.5712}{127, 290--306}

\bibitem[Wong et al.(2004)]{wong-et-al2004}
Wong, M. H., Bjoraker, G. L., Smith, M. D., Flasar, F. M., \& Nixon, C. A.\ 2004. \planss, 52, \href{https://doi.org/10.1016/j.pss.2003.06.005}{385--395}

\bibitem[Wooden et al.(2004)]{wooden-et-al2004}
Wooden, D. H., Woodward, C. E., \& Harker, D. E.\ 2004. \apj, 612, \href{https://doi.org/10.1086/424593}{L77--L80}

\bibitem[Wooden et al.(2017)]{wooden-et-al2017}
Wooden, D. H., Ishii, H. A., \& Zolensky, M. E.\ 2017. Phil. Trans. R. Soc. A., \href{https://doi.org/10.1098/rsta.2016.0260}{375, 20160260}

\bibitem[Yamamoto et al.(2008)]{yamamoto-et-al2008}
Yamamoto, S., Kimura, H., Zubko, E., Kobayashi, H., Wada, K., Ishiguro, M., \& Matsui, T.\ 2008. \apj, \href{https://doi.org/10.1086/527558}{673, L199--L202}

\bibitem[Yanamandra-Fisher \& Hanner(1999)]{yanamandrafisher-hanner1999}
Yanamandra-Fisher, P. A., \& Hanner, M. S.\ 1999. Icarus 138, \href{https://doi.org/10.1006/icar.1998.6066}{107--128}

\end{thebibliography}

{}

\end{document}